\documentclass[preprint,12pt]{elsarticle}

\usepackage[a4paper,margin=1in]{geometry}

\biboptions{numbers,sort&compress}

\usepackage[T1]{fontenc}
\usepackage{amsmath,amssymb}
\usepackage{xspace}
\usepackage{etoolbox}
\usepackage{ragged2e}
\usepackage{siunitx}

\DeclareSIUnit{\angstrom}{\text{\AA}}
\DeclareSIUnit{\mbar}{mbar}

\usepackage{graphicx}
\usepackage{booktabs}
\usepackage{array}
\usepackage{float}
\usepackage{placeins}
\usepackage{pdflscape}
\usepackage{multirow}
\usepackage{microtype}

\usepackage{setspace}
\usepackage{caption}

\usepackage{xcolor}
\usepackage{hyperref}

\hypersetup{
    colorlinks=true,
    linkcolor=blue,
    citecolor=blue,
    urlcolor=blue,
    filecolor=blue,
    pdfborder={0 0 0},
    pdfstartview=FitH
}

\journal{Materials Science and Engineering: A}

\begin{document}

\begin{frontmatter}

% =========================================================
% TITLE
% =========================================================

\title{
Cobalt-Controlled Interphase Partitioning Regulates Matrix Solute
Transport and $\gamma^{\prime}$ Coarsening in Ti-Rich
NiCoCr-Based Superalloys
}

% =========================================================
% AUTHORS
% =========================================================

\author[iisc]{Sudeepta Mukherjee\corref{cor1}}
\ead{sudeeptam@iisc.ac.in}

\author[iisc]{Surendra Kumar Makineni}

\author[iith]{B.~S. Murty}

\author[iisc]{Satyam Suwas\corref{cor1}}
\ead{satyamsuwas@iisc.ac.in}

\cortext[cor1]{
Corresponding authors: Sudeepta Mukherjee and Satyam Suwas.
}

% =========================================================
% AFFILIATIONS
% =========================================================

\address[iisc]{
Department of Materials Engineering,
Indian Institute of Science,
Bengaluru 560012, Karnataka, India
}

\address[iith]{
Department of Materials Science and Metallurgical Engineering,
Indian Institute of Technology Hyderabad,
Kandi, Sangareddy 502285, Telangana, India
}

% =========================================================
% ABSTRACT
% =========================================================

\begin{abstract}
The microstructural stability and mechanical response of $\gamma/\gamma'$ superalloys at elevated temperatures are governed by solute chemistry and elemental partitioning across the $\gamma/\gamma'$ heterophase interface. Conventionally, a lower $\gamma^{\prime}$ solvus temperature and a larger $\gamma/\gamma^{\prime}$ lattice misfit are expected to increase the coarsening rate. Here, Co-for-Ni substitution is used in Ni--Co--Cr--Al--Ti superalloys to systematically modify the $\gamma$-matrix chemistry and generate medium-entropy-alloy-type matrix environments with Ni:Co:Cr ratios of approximately 2:1:2, 2:2:1, and 1:2:1. Despite a $\sim$65~$^{\circ}$C reduction in $\gamma^{\prime}$ solvus temperature, the activation energy for Ti-rich Ni$3$(Al,Ti)-type $L1{2}$ $\gamma^{\prime}$ coarsening increases from $\sim$156 to $\sim$302~kJ~mol$^{-1}$ with increasing Co content. This enhanced coarsening resistance is attributed to an increase in the multicomponent solute-transport resistance, $S=\sum_i (\Delta C_i)^2/(D_i^\gamma C_i^\gamma)$, from $\sim 8.811\times10^{15}$ to $\sim 1.267\times10^{16}$~s~m$^{-2}$, together with a reduction in the apparent interfacial-energy term from 25.0 to 7.55~mJ~m$^{-2}$. The dominant transport resistance correspondingly shifts from Ni/Cr in B--10Co to Ni/Co/Cr in B--30Co.
\end{abstract}
% =========================================================
% KEYWORDS
% =========================================================


\end{frontmatter}

% =========================================================
% MAIN TEXT
% ======================================================
\section{Introduction}

Superalloys derive their exceptional mechanical properties from the presence of $\gamma'$ (Ni$_3$Al) precipitates, which are coherently embedded within the $\gamma$ (face-centered cubic) matrix~\cite{ref1,ref2}. These $\gamma'$ precipitates contribute significantly to high-temperature strength; however, their tendency to coarsen during prolonged thermal exposure at elevated temperatures leads to a gradual degradation of mechanical properties~\cite{ref3,ref4}. The coarsening process, driven by Ostwald ripening, involves the growth of larger precipitates at the expense of smaller ones, ultimately reducing the overall high-temperature performance of the alloy~\cite{ref5}. To mitigate this challenge, the chemical composition of the alloy must be carefully optimized to enhance microstructural stability. Key alloying elements such as aluminum (Al) and titanium (Ti) are well-established for their role in stabilizing $\gamma'$ precipitates, whereas refractory elements such as molybdenum (Mo) and tungsten (W) contribute to solid solution strengthening~\cite{ref6}. In recent years, cobalt (Co) and chromium (Cr) have garnered significant attention for their multifaceted roles in improving $\gamma'$ precipitate stability and overall alloy performance~\cite{ref7}.

Cobalt serves as a $\gamma$ matrix stabilizer and promotes solid solution strengthening in the $\gamma$ channels. It reduces the solubility of aluminium and titanium in the matrix, thereby increasing the $\gamma'$ volume fraction, which enhances the alloy's strength. Additionally, cobalt lowers the stacking fault energy of the alloy~\cite{ref8,ref9}. By reducing the diffusivity of solute elements, Co effectively delays the onset of deleterious phases and extends the thermal stability of $\gamma'$ precipitates even at elevated temperatures~\cite{ref10}. Chromium plays a dual role in superalloys: it enhances oxidation resistance, critical for high-temperature applications, while simultaneously modifying the chemical potential gradients between the $\gamma$ and $\gamma'$ phases, thereby influencing phase stability and reducing coarsening rates~\cite{ref11}. Recent research efforts have been directed toward the development of Ni--Al--Co--Cr--Ti-based superalloys, which offer high strength, excellent thermal stability, and superior resistance to environmental degradation at elevated temperatures~\cite{ref12}. These alloys are designed with objectives such as low cost, reduced density, enhanced mechanical performance, and improved sustainability~\cite{ref13}. Unlike conventional Ni-based superalloys that depend on refractory elements like Mo, W, or Ta for solid solution strengthening, Ni--Al--Co--Cr--Ti alloys exhibit superior thermal stability and mechanical properties even in the absence of these expensive and high-density elements~\cite{ref14,ref15}. This improved performance is attributed to the concentrated nature of the $\gamma$ matrix and the optimized partitioning behavior of Co, Cr, Ti, and Al, which collectively enhance the alloy's long-term high-temperature stability~\cite{ref16,ref17,ref18}.

In this work, the effect of Co variation in the Ni--$x$Co--11Cr--9Al--8Ti ($x=10,20,30$ at.\%) alloy system on the $\gamma'$ coarsening behaviour was systematically investigated. Co-for-Ni substitution was employed as a compositional design strategy to generate distinct medium-entropy-alloy (MEA)-type $\gamma$-matrix environments while maintaining a Ti-rich L1$_2$ $\gamma'$ strengthening phase. Atom probe tomography (APT) was used to quantify the influence of Co on elemental partitioning across the $\gamma/\gamma'$ interface, whereas CALPHAD-based thermokinetic modelling was employed to analyse the diffusion-controlled contribution to $\gamma'$ coarsening. This approach is motivated by the unresolved role of Co in controlling $\gamma'$ coarsening in multicomponent superalloys, with previous studies reporting contradictory effects on $\gamma/\gamma'$ microstructural stability~\cite{ref19,ref20,ref21}.

\section{Materials and Methods}

\subsection{CALPHAD-Guided Alloy Design and Heat Treatment}
CALPHAD-based thermodynamic calculations were performed using Thermo-Calc 2023a with the TCNI11 database \cite{ThermoCalc2022} to guide the design of $\text{Ni--Co--Cr--Al--Ti}$ $\gamma/\gamma'$ superalloys with systematically varied $\gamma$-matrix chemistry. The design objective was to generate distinct $\text{Ni--Co--Cr}$ medium-entropy-alloy-type $\gamma$-matrix environments while retaining a Ti-rich $\text{Ni}_3(\text{Al,Ti})$-type $\text{L1}_2$ $\gamma'$ precipitate phase. Co was therefore substituted at the expense of Ni, while Cr, Al, and Ti were kept nominally fixed. The selected nominal compositions (at.\%) were $\mathrm{Ni}$--11$\mathrm{Cr}$--10$\mathrm{Co}$--9$\mathrm{Al}$--8$\mathrm{Ti}$, $\mathrm{Ni}$--11$\mathrm{Cr}$--20$\mathrm{Co}$--9$\mathrm{Al}$--8$\mathrm{Ti}$, and $\mathrm{Ni}$--11$\mathrm{Cr}$--30$\mathrm{Co}$--9$\mathrm{Al}$--8$\mathrm{Ti}$, hereafter referred to as B--10Co, B--20Co, and B--30Co, respectively.

\begin{figure}[!htbp]
    \centering

    \includegraphics[
        width=0.78\linewidth,
        keepaspectratio
    ]{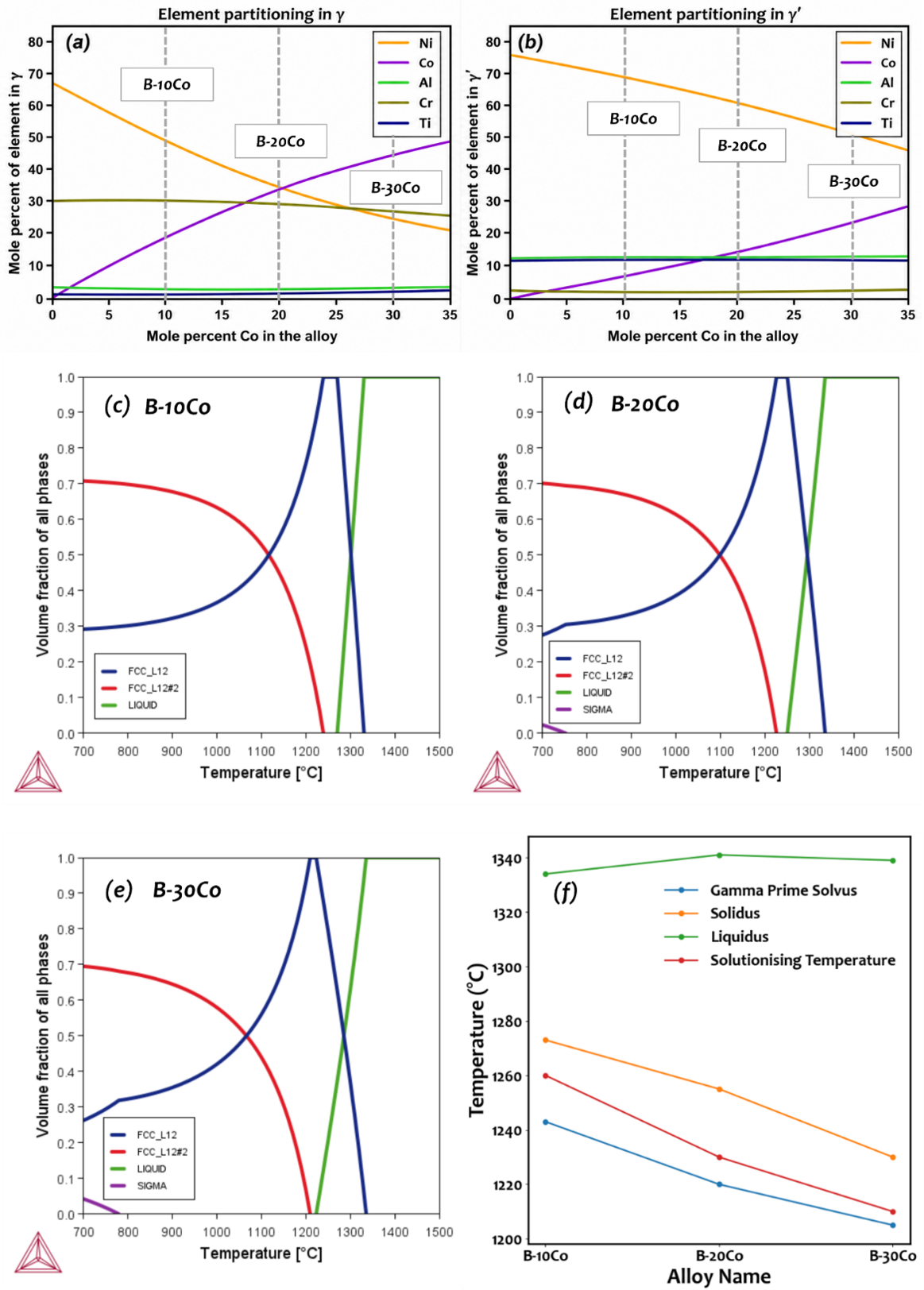}
    \caption{CALPHAD-predicted phase compositions and phase evolution in the designed Ni--Co--Cr--Al--Ti alloys. (a) Calculated compositions of the $\gamma$ and $\gamma^\prime$ phases at \SI{900}{\degreeCelsius} as a function of Co content, showing the evolution of the respective phase chemistries. Calculated phase fractions as a function of temperature for (b1) B--10Co, (b2) B--20Co, and (b3) B--30Co. The calculations confirm the stability of the $\gamma/\gamma^\prime$ two-phase field above \SI{800}{\degreeCelsius}. The predicted transformation temperatures were used to determine the solution-treatment temperatures for the respective alloys. All calculations were performed using Thermo-Calc 2023a with the TCNI11 thermodynamic database. In the Thermo-Calc notation, FCC\_L12\#2 corresponds to the ordered L1$_2$ $\gamma^\prime$ phase, whereas FCC\_L12 corresponds to the disordered FCC $\gamma$ phase.}
    \label{fig:calphad}

\end{figure}

As shown in Figure \ref{fig:calphad}(a), the calculated phase compositions at \SI{900}{\celsius} indicate that increasing Co content progressively enriches the $\gamma$ matrix in Co while reducing its Ni content. Cr remains preferentially partitioned to $\gamma$, whereas Al and Ti remain concentrated in $\gamma'$. Thus, the selected Co additions transform the $\gamma$ matrix from a relatively $\text{Ni--Cr}$-rich environment in B--10Co to a Co-enriched $\text{Ni--Co--Cr}$ medium-entropy matrix in B--30Co without changing the nominal $\gamma'$-forming Al and Ti contents. The alloys were prepared using high-purity elemental constituents in the form of pieces ($\ge 99.9\%$, Alfa Aesar): Ni, Co, Cr, Al, and Ti. Melting was conducted in a vacuum arc furnace on a water-cooled copper hearth under a high-purity (HP, $\sim$99.99\%) Ar atmosphere. For each alloy, a charge of approximately \SI{25}{\gram} was arc-melted, flipped, and re-melted at least five times to promote chemical homogeneity. The buttons were subsequently cast into \SI{3}{\milli\meter} rods.

The phase stability of the designed alloys was evaluated from the calculated phase fractions as a function of temperature, as shown in Figure \ref{fig:calphad}(c-e). The predicted $\gamma'$ solvus decreases with increasing Co content, from approximately \SI{1245}{\celsius} for B--10Co to \SI{1222}{\celsius} for B--20Co and \SI{1184}{\celsius} for B--30Co. Accordingly, the supersolvus homogenisation/solutionising temperatures were selected above the respective $\gamma'$ solvus and below the solidus temperatures (as shown in figure \ref{fig:calphad}(f): \SI{1260}{\celsius} for B--10Co, \SI{1230}{\celsius} for B--20Co, and \SI{1210}{\celsius} for B--30Co. The heat-treatment protocol involved heating to \SI{1175}{\celsius} at \SI{300}{\celsius\per\hour}, followed by a slow ramp of \SI{20}{\celsius\per\hour} to the alloy-specific homogenisation temperatures, holding for \SI{20}{\hour}, and quenching in iced brine. The controlled slow heating through the $\gamma'$ solvus region ensured complete dissolution of precipitates while avoiding incipient melting \cite{Etter2015, Hegde2010}. Post-treatment EPMA-WDS analysis, averaged over 10 measurements, verified chemical homogeneity, with measured compositions within \SI{0.5}{at.\%} of the nominal targets [\ref{tab:S1_wds}].

The calculated phase-fraction evolution further confirms that all three alloys retain a stable $\gamma + \gamma'$ two-phase field at the subsolvus ageing temperatures used for coarsening-kinetics evaluation, namely \SI{900}{\celsius}, \SI{950}{\celsius}, and \SI{1000}{\celsius}. To develop the initial $\gamma/\gamma'$ microstructure, the solutionised rods were vacuum sealed in quartz ampoules under a vacuum of \SI{10e-5}{\mbar} and aged at \SI{900}{\celsius} for \SI{50}{\hour}. For coarsening studies, \SI{4}{\milli\meter}-thick slices of the homogenised alloys were vacuum sealed under the same conditions and isothermally annealed at \SI{900}{\celsius}, \SI{950}{\celsius}, and \SI{1000}{\celsius} for \SI{50}{\hour}, \SI{100}{\hour}, \SI{250}{\hour}, \SI{500}{\hour}, and \SI{1000}{\hour}. After ageing, the samples were quenched again in iced brine to retain the evolved high-temperature microstructure.

\subsection{Thermophysical Property Measurements}
\paragraph{Thermal analysis} for all the samples was performed using a NETZSCH STA 449 F3 Jupiter instrument to evaluate critical phase transformation temperatures, including the $\gamma'$ solvus. Prior to measurement, the sample, weighing approximately 30 mg, was soaked overnight in acetone and subsequently dried. A standardised heating and cooling protocol was followed for all experiments. Approximately \SI{30}{\milli\gram} of each sample was placed in a recrystallised $\text{Al}_2\text{O}_3$ crucible, with an identical crucible used as the reference. The samples were subjected to a complete thermal cycle---heating from \SI{50}{\celsius} to \SI{1300}{\celsius}, followed by cooling to \SI{30}{\celsius}---at a constant rate of \SI{10}{\celsius\per\minute} in both directions. All DSC runs were conducted under a continuous flow of high-purity argon (\SI{60}{\milli\liter\per\minute}) to prevent oxidation. The DSC data were analysed using NETZSCH Proteus software. 

The onset of the first endothermic peak during the heating cycle was identified as the $\gamma'$ solvus temperature. For accurate determination, the method described by Boettinger et al. \cite{Boettinger2006} was adopted, wherein tangents were drawn on the flanks of the asymptotic peak, and their point of intersection was used to define the onset temperature with precision.

\subsection{Microstructural Characterisation and Analysis}
\textbf{X-ray diffraction (XRD)} was used to determine the $\gamma/\gamma'$ lattice misfit in the peak-aged condition. Measurements were carried out using a Rigaku SmartLab high-resolution diffractometer with $\text{Cu K}\alpha$ radiation, having a wavelength $\lambda = \SI{0.15406}{\nano\meter}$, operated at \SI{40}{\kilo\volt} and \SI{40}{\milli\ampere}. The diffractometer was equipped with Johansson optics to suppress $\text{K}\alpha_2$ and $\text{K}\beta$ radiation, thereby improving angular resolution and peak-position accuracy.
For $\gamma/\gamma'$ lattice-misfit determination, high-resolution scans were collected over a $2\theta$ range of \SI{49}{\degree}--\SI{53}{\degree} using a step size of \SI{0.008}{\degree}, covering the closely spaced $\gamma$ and $\gamma'$ reflections. The overlapping diffraction peaks were deconvoluted using pseudo-Voigt functions to accurately resolve the individual $\gamma$ and $\gamma'$ peak positions. The corresponding lattice parameters of the $\gamma$ matrix and $\gamma'$ precipitate phases were then calculated from the fitted peak positions. The $\gamma/\gamma'$ lattice misfit, $\delta$, was determined using Equation~\ref{eq:misfit}:
\begin{equation}
    \delta = \frac{2(a_{\gamma'} - a_{\gamma})}{a_{\gamma'} + a_{\gamma}}
    \label{eq:misfit}
\end{equation}
where $a_{\gamma}$ and $a_{\gamma'}$ are the lattice parameters of the $\gamma$ and $\gamma'$ phases, respectively.

\textbf{Backscattered electron (BSE) imaging} was performed using a ZEISS GeminiSEM 450 at \SI{30}{\kilo\volt}/\SI{13}{\nano\ampere}, providing contrast between the $\gamma$ and $\gamma'$ phases. Quantitative elemental analysis was carried out using JEOL JXA-8530F EPMA with WDS mode, at \SI{15}{\kilo\volt} and \SI{50}{\nano\ampere}. The measurements were averaged across multiple regions for accuracy (See Supplementary Table S1).

\textbf{Volume fractions of $\gamma'$ precipitates} of the heat-treated alloy were measured using the grid method (ASTM standard E562-11) \cite{ASTM_E562} using a Python-based script. The measurements were carried out utilizing 2-D projected SEM images of near $[100]$ oriented grains for at least 200 precipitates. The precipitate size calculation was based on considering circular area-equivalent radius. The area of individual precipitates ($A$) was traced from the SEM images and from which the circular area equivalent radius, $r$, (precipitate size) was calculated by Equation~\ref{eq:radius}:
\begin{equation}
    r = \sqrt{\frac{A}{\pi}}
    \label{eq:radius}
\end{equation}

The error ($s$) in the measurement of average precipitate size was calculated by Equation~\ref{eq:error}:
\begin{equation}
    s = \frac{\sigma}{\sqrt{N}}
    \label{eq:error}
\end{equation}
where $\sigma$ is the standard deviation of the average precipitate size and $N$ is the number of precipitates used for the measurement.

\subsection{Atom Probe Tomography}
Specimens for atomic-scale compositional measurements were prepared from polished, peak-aged B-10Co, B-20Co, and B-30Co samples heat-treated at \SI{900}{\degreeCelsius} for \SI{50}{\hour}. Site-specific needle-shaped specimens were extracted using a dual-beam SEM/focused-ion-beam instrument (FIB, Helios Nanolab 600) following a standard in-situ lift-out protocol~\cite{Thompson_fibref1,Makineni_fibref2}. Protective platinum (Pt) layers were deposited prior to milling to prevent surface damage to the regions of interest. The tips were sharpened by multi-step gallium (Ga) ion milling at \SI{30}{\kilo\volt}, with progressively reduced ion-beam currents, followed by final low-energy cleaning at \SI{2}{\kilo\volt} and \SI{8}{\pico\ampere} to minimise FIB-induced damage and remove the severely damaged surface regions produced by the Ga ion source. The final tip radii were approximately $\sim\SI{50}{\nano\meter}$. 
 Atomic-scale compositional analysis across the $\gamma/\gamma'$ interfaces was performed using a local electrode atom probe, LEAP\texttrademark~5000XR (Cameca Instruments Inc.), operated in laser-pulsing mode. Ultraviolet picosecond laser pulses were applied at a repetition rate of \SI{125}{\kilo\hertz}, with a laser pulse energy of \SI{55}{\pico\joule}. The target detection rate was maintained at $0.5\%$, and the specimen base temperature was kept at \SI{70}{\kelvin} during analysis.
 Data reconstruction and preliminary compositional analysis were carried out using IVAS\texttrademark~3.8.10 software. 

\section{Results}

\subsection{Microstructure after ageing at \SI{900}{\celsius} for \SI{50}{\hour}}

\begin{figure}[!htbp]
\centering
\includegraphics[width=0.9\linewidth,keepaspectratio]{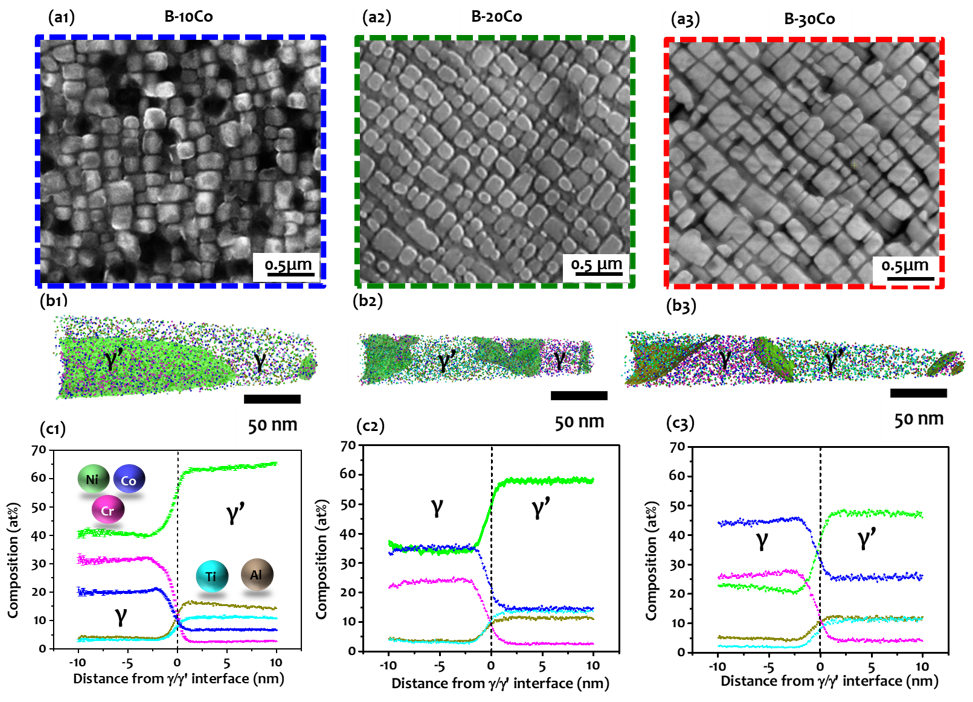}
\caption{Back-scattered-electron (BSE) SEM images of (a1) B--10Co, (a2) B--20Co, and (a3) B--30Co alloys showing the $\gamma/\gamma^{\prime}$ microstructure. APT reconstructions showing the distribution of Al, Ti, Cr, Ni, and Co across the $\gamma/\gamma^{\prime}$ microstructure are shown for (b1) B--10Co, (b2) B--20Co, and (b3) B--30Co. The corresponding elemental proxigrams in (c1--c3) illustrate the compositional profiles across the $\gamma/\gamma^{\prime}$ interfaces, highlighting the elemental partitioning behaviour in the respective alloys.}
\label{fig:microstructure_sem}
\end{figure}

Figure~\ref{fig:microstructure_sem}(a1--a3) shows back-scattered-electron (BSE) SEM images of the B--10Co, B--20Co, and B--30Co alloys after ageing at \SI{900}{\celsius} for \SI{50}{\hour}. All three alloys exhibit a two-phase $\gamma/\gamma^{\prime}$ microstructure, with systematic changes in the size, volume fraction, and morphology of the $\gamma^{\prime}$ precipitates as a function of Co content. With increasing Co content, the $\gamma^{\prime}$ precipitates become progressively finer. Quantitative stereological measurements reveal a decrease in the average $\gamma^{\prime}$ precipitate radius from $\sim\SI{106}{\nano\meter}$ for B--10Co to $\sim\SI{82}{\nano\meter}$ for B--20Co and $\sim\SI{75}{\nano\meter}$ for B--30Co, accompanied by a decrease in the $\gamma^{\prime}$ volume fraction. Supplementary Figure~S1(i) shows the corresponding low-magnification images used for the precipitate-size-distribution analysis presented in Supplementary Figure~S3(i).

Precipitate morphology was further evaluated using the shape factor, $\chi = 4\pi A/P^2$, where $A$ is the precipitate area and $P$ is the precipitate perimeter \cite{Rakoczy2020}. Values of $\chi$ close to 1 indicate a nearly spherical morphology, whereas values in the range of 0.8--0.9 indicate rounded-cuboidal precipitates. The estimated shape factor values are 0.85 for B--10Co, 0.84 for B--20Co, and 0.83 for B--30Co, indicating that the precipitates retain a rounded-cuboidal morphology with increasing Co content. This suggests that Co addition reduces the precipitate size while maintaining shape uniformity. The measured density of all three alloys is approximately \SI{7.44}{\gram\per\centi\meter\cubed}, which is lower than that of conventional Ni-based and Co-based superalloys \cite{Elliott2005}.

\noindent\textbf{Atomic-scale compositional analysis:}
To accurately quantify elemental partitioning between the $\gamma$ and $\gamma^{\prime}$ phases, atom probe tomography (APT) experiments were conducted on B--10Co, B--20Co, and B--30Co alloys aged at \SI{900}{\celsius} for \SI{50}{\hour}. Figure~\ref{fig:microstructure_sem}(b1--b3) shows the APT reconstructions for the three alloys, displaying the spatial distribution of Ni, Co, Cr, Ti, and Al atoms.

\begin{figure}[!htbp]
\centering
\includegraphics[width=0.9\linewidth,keepaspectratio]{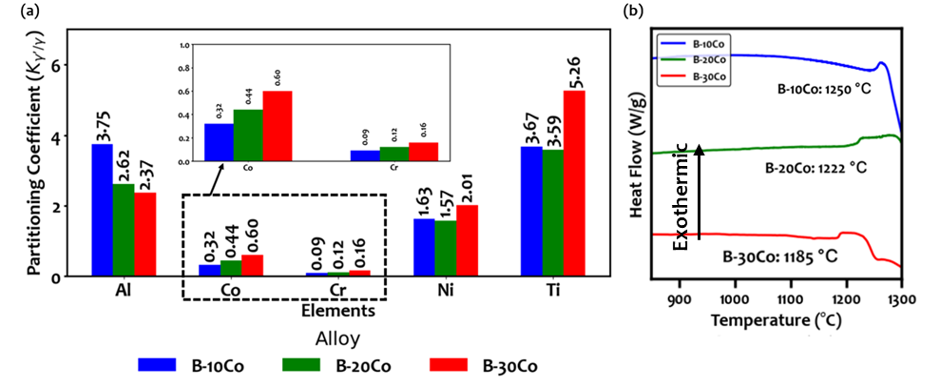}
\caption{(a) Comparison of solute partitioning coefficients across the $\gamma/\gamma^{\prime}$ interfaces in B--10Co, B--20Co, and B--30Co alloys. (b) Differential scanning calorimetry (DSC) heating curves for the aged alloys, showing the decrease in the $\gamma^{\prime}$ solvus temperature with increasing Co content.}
\label{fig:partitioning_dsc}
\end{figure}

 The $\gamma/\gamma^{\prime}$ interface was identified using iso-composition surfaces of \SI{50}{at.\%} Ni for B--10Co, \SI{45}{at.\%} Ni for B--20Co, and \SI{35}{at.\%} Ni for B--30Co. The $\gamma^{\prime}$ precipitates are surrounded by $\gamma$ channels, with the phase distinction made based on preferential solute partitioning: Ni, Al, and Ti partition preferentially to $\gamma^{\prime}$, whereas Co and Cr partition preferentially to the $\gamma$ matrix \cite{Mukhopadhyay2021,Nithin2017}. The comparison of solute partitioning coefficients across the $\gamma/\gamma^{\prime}$ interfaces is shown in Figure~\ref{fig:partitioning_dsc}(a), while the corresponding measured phase compositions and partitioning coefficients are listed in Table~\ref{tab:measured_compositions}.

The composition of $\gamma^{\prime}$ indicates a stoichiometry close to $\text{(Co,Ni,Cr)}_3\text{(Al,Ti)}$, consistent with an A$_3$B-type L1$_2$ crystal structure. In this structure, Co, Ni, and Cr occupy the A sublattice positions, while Al and Ti occupy the B sublattice positions \cite{Pandey2019}. Specifically, the $(\text{Ni}+\text{Co}+\text{Cr})/(\text{Al}+\text{Ti})$ ratio in the $\gamma^{\prime}$ phase is close to $\sim3$ for all alloys. This ratio is further used to estimate the approximate sublattice occupancy of elements and to infer the effect of phase chemistry on the thermophysical properties, as discussed in Section~4.1.

\begin{table}[!htb]
\centering
\caption{Measured compositions of $\gamma$ and $\gamma^{\prime}$ phases in the B--10Co, B--20Co, and B--30Co alloys after ageing at \SI{900}{\celsius} for \SI{50}{\hour}, together with the corresponding partitioning coefficients, $K=C_i^{\gamma^{\prime}}/C_i^{\gamma}$.}
\label{tab:measured_compositions}
\resizebox{0.82\linewidth}{!}{
\begin{tabular}{l c c c c c}
\toprule
\textbf{Alloy / Phase} & \textbf{Ni (at.\%)} & \textbf{Cr (at.\%)} & \textbf{Co (at.\%)} & \textbf{Al (at.\%)} & \textbf{Ti (at.\%)} \\
\midrule
B--10Co ($\gamma$) & $40.2 \pm 1.2$ & $31.2 \pm 1.1$ & $20.8 \pm 1.0$ & $3.8 \pm 0.5$ & $2.9 \pm 0.4$ \\
B--10Co ($\gamma^{\prime}$) & $65.5 \pm 0.3$ & $2.7 \pm 0.1$ & $6.6 \pm 0.2$ & $14.3 \pm 0.2$ & $10.7 \pm 0.2$ \\
B--10Co ($K \pm \sigma$) & $1.6 \pm 0.1$ & $0.1 \pm 0.1$ & $0.3 \pm 0.1$ & $3.8 \pm 0.5$ & $3.7 \pm 0.5$ \\
\midrule
B--20Co ($\gamma$) & $37.0 \pm 0.6$ & $21.6 \pm 0.5$ & $33.4 \pm 0.5$ & $4.2 \pm 0.2$ & $3.8 \pm 0.2$ \\
B--20Co ($\gamma^{\prime}$) & $58.3 \pm 0.5$ & $2.5 \pm 0.2$ & $14.6 \pm 0.4$ & $11.0 \pm 0.3$ & $13.6 \pm 0.4$ \\
B--20Co ($K \pm \sigma$) & $1.6 \pm 0.1$ & $0.1 \pm 0.1$ & $0.4 \pm 0.1$ & $2.6 \pm 0.2$ & $3.6 \pm 0.2$ \\
\midrule
B--30Co ($\gamma$) & $22.9 \pm 0.4$ & $25.7 \pm 0.5$ & $44.2 \pm 0.5$ & $5.1 \pm 0.2$ & $2.2 \pm 0.2$ \\
B--30Co ($\gamma^{\prime}$) & $46.0 \pm 0.7$ & $4.0 \pm 0.3$ & $26.3 \pm 0.6$ & $12.1 \pm 0.5$ & $11.6 \pm 0.5$ \\
B--30Co ($K \pm \sigma$) & $2.0 \pm 0.1$ & $0.2 \pm 0.1$ & $0.6 \pm 0.1$ & $2.4 \pm 0.1$ & $5.3 \pm 0.2$ \\
\bottomrule
\end{tabular}
}
\end{table}

\begin{table}[!htb]
\centering
\caption{Approximate elemental ratios describing the chemistry of the $\gamma$ matrix and $\gamma^{\prime}$ precipitates determined from APT measurements.}
\label{tab:elemental_ratios}

\small
\renewcommand{\arraystretch}{1.25}

\begin{tabular}{l c c c}
\toprule
\textbf{Alloy} &
\shortstack{\textbf{Approx. Ni:Co:Cr}\\\textbf{ratio in $\gamma$}} &
\shortstack{\textbf{Approx. Ni:Co}\\\textbf{ratio in $\gamma^{\prime}$}} &
\shortstack{\textbf{Approx. Al:Ti}\\\textbf{ratio in $\gamma^{\prime}$}} \\
\midrule
B--10Co & 2:1:2 & 2:0.2 & 1.3:1 \\
B--20Co & 2:2:1 & 2:0.5 & 0.8:1 \\
B--30Co & 1:2:1 & 2:1.1 & 1.0:1 \\
\bottomrule
\end{tabular}

\end{table}

\noindent\textbf{Volume fraction of phases as a function of nominal Co content:}
Stereological analysis of the microstructure images reveals a decrease in the $\gamma^{\prime}$ volume fraction from 0.753 for B--10Co to 0.635 for B--30Co. This decreasing trend is consistent with the $\gamma^{\prime}$ volume fraction estimated from APT composition measurements using the lever rule. Since the microstructure contains only $\gamma$ and $\gamma^{\prime}$ phases with similar densities, the $\gamma^{\prime}$ volume fraction, $f_{\gamma^{\prime}}$, was estimated using:

\begin{equation}
f_{\gamma^{\prime}} =
\frac{C_i^{\mathrm{alloy}} - C_i^{\gamma}}
{C_i^{\gamma^{\prime}} - C_i^{\gamma}}
\label{eq:lever_rule}
\end{equation}

\begin{figure}[!htbp]
\centering
\includegraphics[width=0.9\linewidth,keepaspectratio]{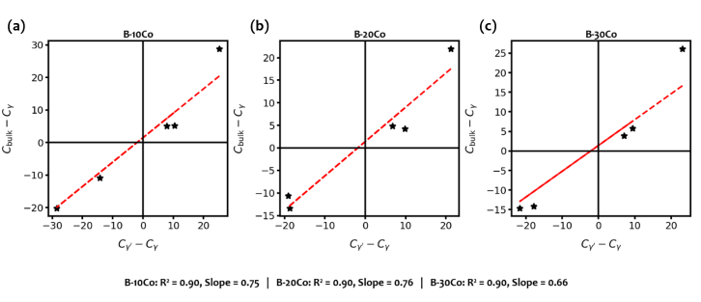}
\caption{Plot of $(C_i^{\mathrm{alloy}} - C_i^{\gamma})$ against $(C_i^{\gamma^{\prime}} - C_i^{\gamma})$ used to determine the $\gamma^{\prime}$ volume fraction in (a) B--10Co, (b) B--20Co, and (c) B--30Co alloys via the lever rule.}
\label{fig:vf}
\end{figure}

The regression slopes yield direct estimates of the $\gamma^{\prime}$ volume fractions, as shown in Figure~\ref{fig:vf}. The decrease in $\gamma^{\prime}$ volume fraction with increasing Co content can be attributed primarily to the reduction in Al partitioning to the $\gamma^{\prime}$ phase.

\subsection{Comparison of $\gamma/\gamma^{\prime}$ lattice misfit as a function of Co content}

\begin{figure}[!htbp]
\centering
\includegraphics[width=0.9\linewidth,keepaspectratio]{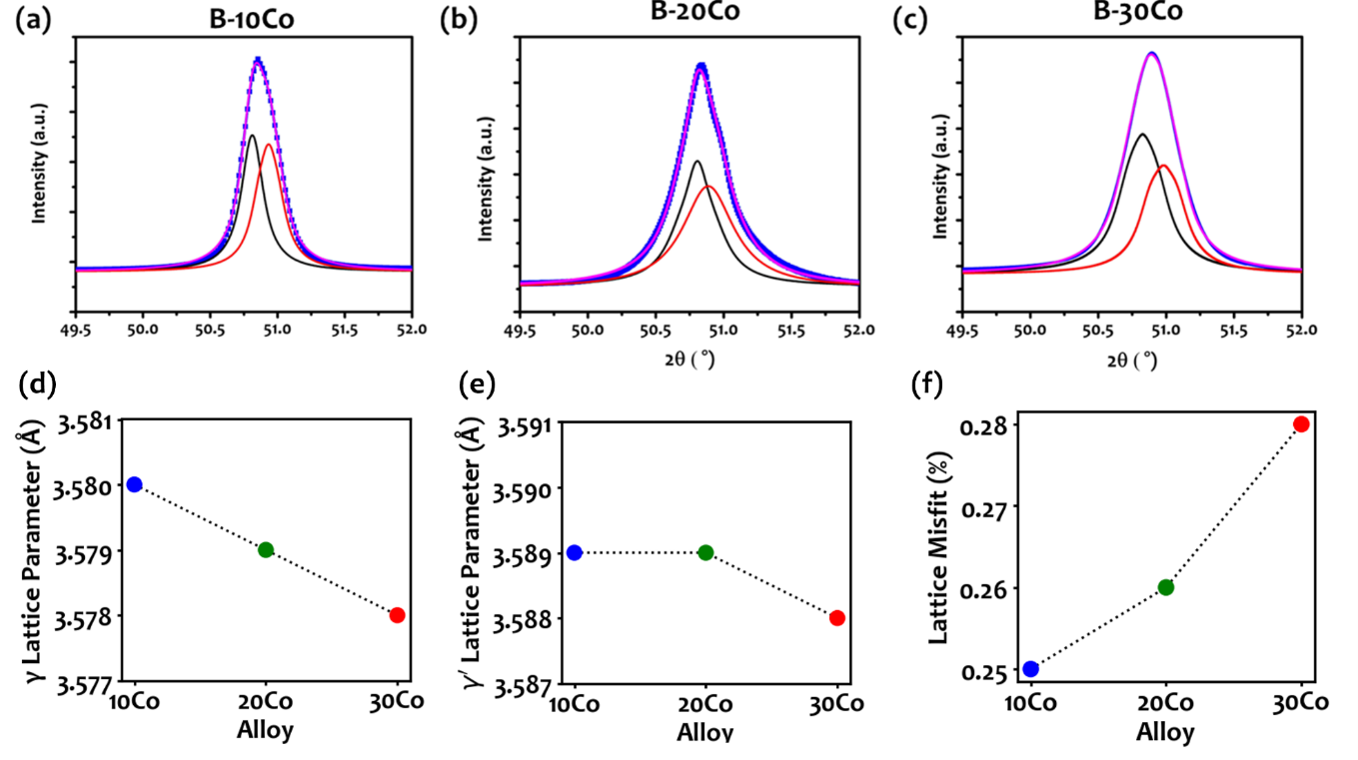}
\caption{X-ray diffraction (XRD) analysis of (a) B--10Co, (b) B--20Co, and (c) B--30Co alloys after ageing at \SI{900}{\celsius} for \SI{50}{\hour}. Deconvolution of the (200) peak intensity using pseudo-Voigt functions reveals two contributions from the $\gamma$ and $\gamma^{\prime}$ phases.}
\label{fig:xrd_analysis}
\end{figure}

Figure~\ref{fig:xrd_analysis} shows the X-ray diffraction peak of the (200) reflection from all three alloys. The peaks were deconvoluted using a pseudo-Voigt function. The $\gamma$ and $\gamma^{\prime}$ peaks were distinguished based on their relative peak intensities and full width at half maximum. The broader and more intense peak was attributed to $\gamma^{\prime}$, consistent with its higher volume fraction and finer size distribution in the microstructure. The calculated constrained $\gamma/\gamma^{\prime}$ lattice misfit values for the B--10Co, B--20Co, and B--30Co alloys, based on the extracted lattice parameters, are summarized in Table~\ref{tab:lattice_misfit} and plotted in Figure~\ref{fig:microstructure_sem}(d--f).

The misfit values for all three alloys are positive, unlike most commercial Ni-based superalloys \cite{Mughrabi2014}, which typically exhibit negative misfit. These alloys therefore resemble Co-based superalloys in terms of the sign of misfit, indicating a larger $\gamma^{\prime}$ unit cell relative to the $\gamma$ matrix \cite{Makineni2021}. Lattice misfit depends on elemental partitioning across the $\gamma/\gamma^{\prime}$ interface and on the site occupancy of solutes in the ordered L1$_2$ $\gamma^{\prime}$ unit cell.

\begin{table}[!htb]
\centering
\caption{Lattice parameters and constrained lattice misfits derived from XRD measurements for B--10Co, B--20Co, and B--30Co alloys.}
\label{tab:lattice_misfit}
\begin{tabular}{l c c c}
\toprule
\textbf{Alloy} & \textbf{$\boldsymbol{\gamma^{\prime}}$ lattice parameter (\si{\angstrom})} & \textbf{$\boldsymbol{\gamma}$ lattice parameter (\si{\angstrom})} & \textbf{Lattice misfit (\%)} \\
\midrule
B--10Co & 3.582 & 3.572 & 0.25 \\
B--20Co & 3.583 & 3.574 & 0.26 \\
B--30Co & 3.585 & 3.565 & 0.28 \\
\bottomrule
\end{tabular}
\end{table}

\subsection{Coarsening kinetics of $\gamma^{\prime}$ precipitates as a function of Co content}

\begin{figure}[!htbp]
\centering
\includegraphics[width=0.78\linewidth,keepaspectratio]{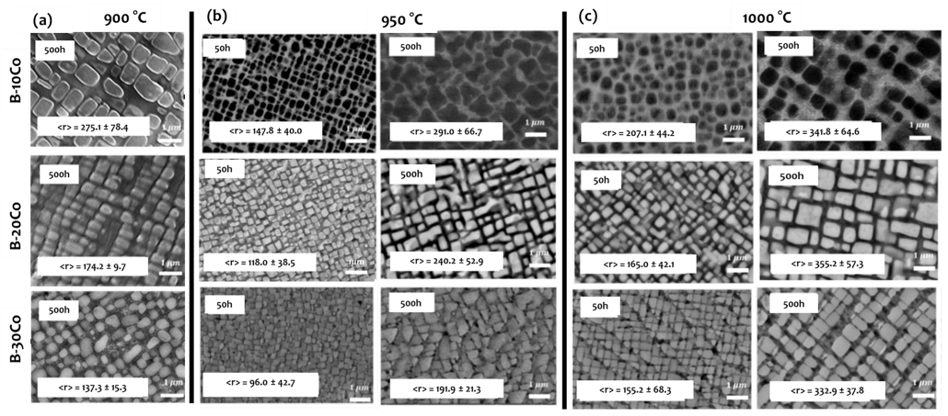}

\vspace{2pt}

\includegraphics[width=0.58\linewidth,keepaspectratio]{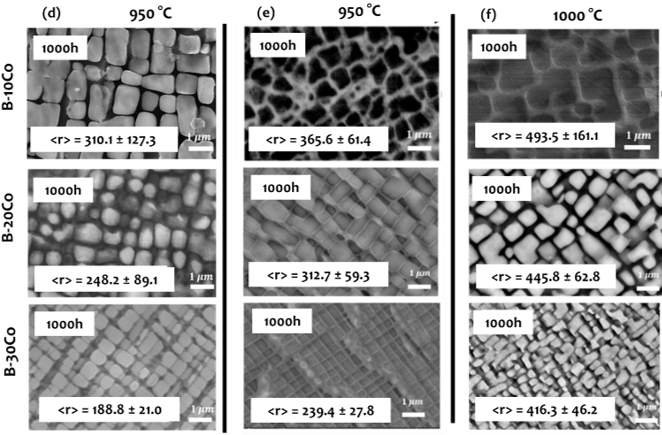}
\caption{Representative SEM micrographs of the B--10Co, B--20Co, and B--30Co alloys isothermally heat-treated at \SI{900}{\celsius}, \SI{950}{\celsius}, and \SI{1000}{\celsius}, showing microstructural evolution during isothermal holding for 50, 500, and \SI{1000}{\hour}.}
\label{fig:sem_evolution}
\end{figure}

The temporal evolution of $\gamma^{\prime}$ precipitates in B--10Co, B--20Co, and B--30Co alloys was investigated during isothermal annealing at \SI{900}{\celsius}, \SI{950}{\celsius}, and \SI{1000}{\celsius}. Figure~\ref{fig:sem_evolution} presents representative BSE SEM images for the alloys annealed at these temperatures for selected durations.

\subsubsection{Microstructures after isothermal annealing at \SI{900}{\celsius}}

The microstructures show that the $\gamma^{\prime}$ precipitates are uniformly distributed with a rounded-cuboidal morphology up to \SI{1000}{\hour} in all three alloys. There is no indication of directional coarsening, rafting, or coalescence within the grains or at grain boundaries, indicating good microstructural stability. Even after \SI{1000}{\hour}, similar morphology persists, with no evidence of secondary phases such as topologically close-packed phases. Quantitative stereological analysis shows that the mean $\gamma^{\prime}$ precipitate radius, $\langle r \rangle$, increases with time, consistent with Ostwald ripening.

\subsubsection{Microstructure after isothermal annealing at \SI{950}{\celsius} and \SI{1000}{\celsius}}

At both temperatures, the $\gamma^{\prime}$ precipitates retained a rounded-cuboidal morphology. Precipitate coarsening was evident from the progressive increase in $\langle r \rangle$ with time. After \SI{1000}{\hour} of ageing, the mean precipitate sizes were $366 \pm \SI{61}{\nano\meter}$ for B--10Co, $313 \pm \SI{59}{\nano\meter}$ for B--20Co, and $239 \pm \SI{29}{\nano\meter}$ for B--30Co at \SI{950}{\celsius}; and $494 \pm \SI{16}{\nano\meter}$ for B--10Co, $446 \pm \SI{6}{\nano\meter}$ for B--20Co, and $416 \pm \SI{43}{\nano\meter}$ for B--30Co at \SI{1000}{\celsius}. These values indicate slower coarsening with increasing Co content.

\subsubsection{Comparison of evolution of $\langle r \rangle$ as a function of time and Co content}

To isolate coarsening from concurrent nucleation and growth processes, all samples were peak-aged prior to extended exposures. Log--log plots of average precipitate radius, $\langle r \rangle$, versus time were constructed at each temperature to assess coarsening kinetics under thermodynamically stable conditions.

To isolate coarsening from concurrent nucleation and growth processes, all samples were peak-aged prior to extended exposures. Log--log plots of average precipitate radius, $\langle r \rangle$, versus ageing time were constructed at each temperature to assess coarsening kinetics under thermodynamically stable conditions.

\begin{figure}[!htbp]
\centering
\includegraphics[width=0.78\linewidth,keepaspectratio]{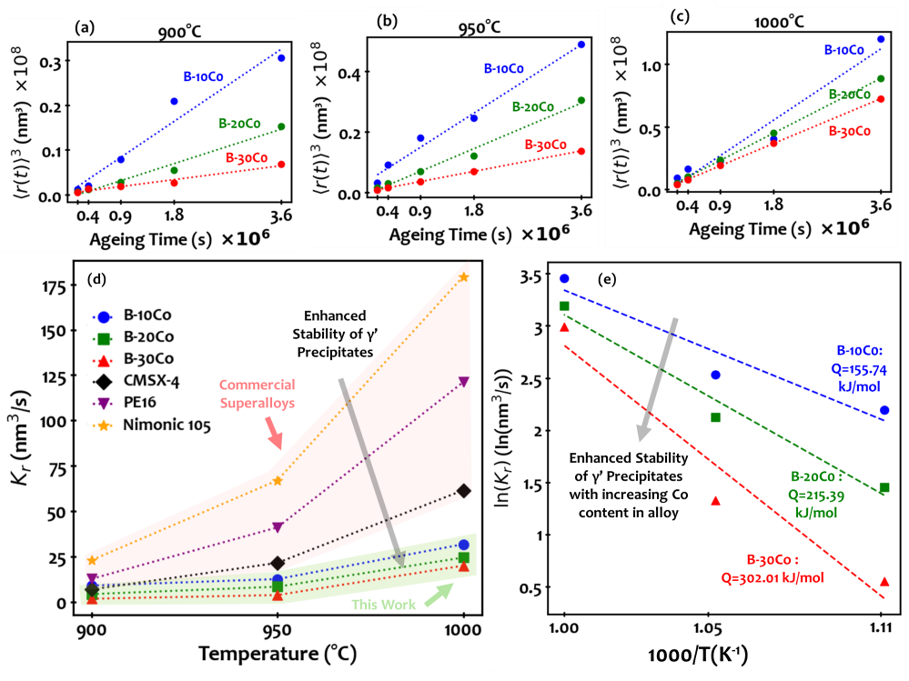}
\caption{Temporal evolution of $\gamma^{\prime}$ precipitates in B--10Co, B--20Co, and B--30Co alloys during isothermal ageing at (a) \SI{900}{\celsius}, (b) \SI{950}{\celsius}, and (c) \SI{1000}{\celsius}. The plots of $\langle r(t) \rangle^3$ versus ageing time demonstrate a linear trend indicative of diffusion-controlled coarsening. The slope of each linear fit provides the coarsening rate constant, $K_r$. (d) Summary of coarsening rate constants for B--xCo alloys compared with selected commercial superalloys and back-calculated $K_r$ values at \SI{900}{\celsius}, \SI{950}{\celsius}, and \SI{1000}{\celsius} \cite{Lapin2008}. (e) Arrhenius plot of $\ln(K_r)$ versus $1000/T$ used to extract the activation energy for $\gamma^{\prime}$ coarsening.}
\label{fig:coarsening_plots}
\end{figure}

\begin{table}[!htb]
\centering
\caption{Coarsening rate constants, $K_r$, at \SI{900}{\celsius}, \SI{950}{\celsius}, and \SI{1000}{\celsius}, and activation energies for $\gamma^{\prime}$ coarsening in B--10Co, B--20Co, and B--30Co alloys derived from $\langle r(t) \rangle^3$ versus ageing-time plots and Arrhenius-type analysis.}
\label{tab:coarsening_rate_constants}
\resizebox{0.8\linewidth}{!}{
\begin{tabular}{l c c c c}
\toprule
\textbf{Alloy} &
\textbf{Activation energy, $\boldsymbol{Q}$} &
\textbf{$\boldsymbol{K_r}$ at \SI{900}{\celsius}} &
\textbf{$\boldsymbol{K_r}$ at \SI{950}{\celsius}} &
\textbf{$\boldsymbol{K_r}$ at \SI{1000}{\celsius}} \\
&
\textbf{(\si{\kilo\joule\per\mol})} &
\textbf{(\si{\nano\meter\cubed\per\second})} &
\textbf{(\si{\nano\meter\cubed\per\second})} &
\textbf{(\si{\nano\meter\cubed\per\second})} \\
\midrule
B--10Co & 155.74 & 8.98 & 12.57 & 31.75 \\
B--20Co & 215.39 & 4.28 & 8.35  & 24.41 \\
B--30Co & 302.01 & 1.73 & 3.76  & 19.96 \\
\bottomrule
\end{tabular}
}
\end{table}

Supplementary Figure~S3 shows the temporal evolution of $\langle r \rangle$ at \SI{900}{\celsius}, \SI{950}{\celsius}, and \SI{1000}{\celsius}, respectively, by plotting $\log \langle r \rangle$ against $\log t$. The corresponding $\langle r \rangle^3$ versus $t$ plots are shown in Figure~\ref{fig:coarsening_plots}. The calculated temporal exponents and coarsening rate constants indicate matrix-diffusion-controlled coarsening kinetics \cite{Pandey2021}. At \SI{900}{\celsius}, $K_r$ decreases from \SI{8.98}{\nano\meter\cubed\per\second} for B--10Co to \SI{4.28}{\nano\meter\cubed\per\second} for B--20Co and \SI{1.73}{\nano\meter\cubed\per\second} for B--30Co. A similar trend is observed at \SI{950}{\celsius} and \SI{1000}{\celsius}, as summarized in Table~\ref{tab:coarsening_rate_constants}. These results indicate that coarsening accelerates with increasing temperature for all three alloys, while increasing Co content systematically reduces the coarsening rate at each ageing temperature.

The coarsening process is thermally activated and follows an Arrhenius-type relationship between the coarsening rate constant, $K_r$, and absolute temperature, $T$, expressed as \cite{Li2009}:

\begin{equation}
K_r = K_0 \exp\left(-\frac{Q}{RT}\right)
\label{eq:arrhenius}
\end{equation}

where $K_0$ is the pre-exponential factor, $Q$ is the activation energy for coarsening, $R$ is the universal gas constant, and $T$ is the absolute temperature in Kelvin. By plotting $\ln(K_r)$ against $1/T$, the activation energy can be extracted from the slope of the fitted linear curve, as shown in Figure~\ref{fig:coarsening_plots}(e).

The calculated activation energies for coarsening are \SI{155.74}{\kilo\joule\per\mol} for B--10Co, \SI{215.39}{\kilo\joule\per\mol} for B--20Co, and \SI{302.01}{\kilo\joule\per\mol} for B--30Co. The activation energy for B--30Co is therefore nearly twice that of B--10Co, indicating a substantially higher energy barrier for solute redistribution during $\gamma^{\prime}$ coarsening. Since the coarsening behaviour is primarily governed by matrix diffusion, the higher activation energy in B--30Co suggests that Co enrichment of the $\gamma$ matrix slows solute transport and enhances the thermal stability of the $\gamma^{\prime}$ precipitates.
\section{Discussion}
\subsection{Effect of Co partitioning on $\gamma/\gamma^{\prime}$ phase chemistry and $\gamma^{\prime}$ solvus temperature}

A reduction in the $\gamma^{\prime}$ solvus temperature is observed with increasing Co content, decreasing from $\sim\SI{1250}{\degreeCelsius}$ for B--10Co to $\sim\SI{1185}{\degreeCelsius}$ for B--30Co. This reduction cannot be explained solely by the nominal Co-for-Ni substitution, but is closely linked to the accompanying changes in $\gamma/\gamma^{\prime}$ elemental partitioning and the resulting modification of the L1$_2$ $\gamma^{\prime}$ chemistry. With increasing Co content, the partitioning coefficient of Al decreases from $K_{\mathrm{Al}}\sim3.8$ to $\sim2.4$, whereas the partitioning coefficients of Ti, Ni, and Co increase from $K_{\mathrm{Ti}}\sim3.6$ to $\sim5.3$, $K_{\mathrm{Ni}}\sim1.6$ to $\sim2.0$, and $K_{\mathrm{Co}}\sim0.3$ to $\sim0.6$, respectively. This indicates that the $\gamma^{\prime}$ phase becomes progressively enriched in Ti, Ni, and Co, while the relative Al enrichment in $\gamma^{\prime}$ decreases.

The decrease in $\gamma^{\prime}$ solvus temperature with increasing Co content can therefore be rationalized by the combined effect of three factors. First, the reduction in Al partitioning weakens the conventional Ni$_3$Al-type L1$_2$ chemistry, which is primarily responsible for the high $\gamma^{\prime}$ ordering stability in Ni-based superalloys. Second, the increased enrichment of Co and Ti in $\gamma^{\prime}$ enhances the probability of Co--Ti and Ni--Ti bonding within the L1$_2$ unit cell. Although Ni$_3$Ti does not crystallize in the L1$_2$ structure in the binary Ni--Ti system, the Co--Ti binary system contains a stable L1$_2$ Co$_3$Ti phase with a considerably lower solvus temperature of $\sim\SI{1005}{\degreeCelsius}$ \cite{massalski1990,im2018elemental}. Third, increased Co and Cr contents are known to reduce the ordering energy of the L1$_2$ Ni$_3$Al structure \cite{zhu2023abinitio}. Hence, the observed reduction in $\gamma^{\prime}$ solvus temperature is attributed to the combined effects of reduced Al partitioning, increased Co--Ti/Ni--Ti bonding tendency, and reduced L1$_2$ ordering stability.

The Co-induced changes in phase chemistry also influence the $\gamma/\gamma^{\prime}$ lattice parameters and lattice misfit. Both $\gamma$ and $\gamma^{\prime}$ lattice parameters show a marginal decrease with increasing Co content. However, the positive $\gamma/\gamma^{\prime}$ lattice misfit increases from $\sim0.25\%$ in B--10Co to $\sim0.28\%$ in B--30Co, indicating that the contraction of the $\gamma$ matrix is relatively greater than that of the $\gamma^{\prime}$ phase. This trend can be understood from the APT-derived elemental partitioning. With increasing Co content, the $\gamma$ matrix is progressively enriched in Co and depleted in Ni, while Ti shows a substantial reduction in the $\gamma$ phase, particularly from B--20Co to B--30Co. Since Ti has a relatively large atomic radius of $\sim\SI{147}{\pico\meter}$, its depletion from the $\gamma$ matrix contributes significantly to the contraction of the $\gamma$ lattice parameter.

In contrast, the lattice parameter of the $\gamma^{\prime}$ phase is controlled by competing effects. Based on the measured phase compositions, Co is inferred to substitute primarily on the Ni sublattice, whereas Ti substitutes on the Al sublattice of the L1$_2$ structure. This substitution increases the fraction of Co--Ti bonds in $\gamma^{\prime}$. The L1$_2$ Co$_3$Ti phase has a larger lattice parameter of $\sim\SI{3.606}{\angstrom}$ compared with $\sim\SI{3.570}{\angstrom}$ for Ni$_3$Al \cite{massalski1990,im2018elemental}. In addition, first-principles calculations have shown that Co--Ni and Ti--Ni bonding environments in L1$_2$ Co--Ti-containing alloys can introduce stronger and more directional bonding than the native Ni--Al bonds, generating additional steric strain and promoting lattice expansion \cite{gong2023alloying}. However, this expansion tendency is partly offset by the substantial reduction in Ni content in $\gamma^{\prime}$, from $\sim65$ at.\% in B--10Co to $\sim46$ at.\% in B--30Co. As a result, the $\gamma^{\prime}$ lattice parameter decreases only marginally, whereas the $\gamma$ matrix contracts more strongly. The relatively greater contraction of $\gamma$ compared with $\gamma^{\prime}$ therefore leads to the observed increase in positive $\gamma/\gamma^{\prime}$ lattice misfit with increasing Co content.

\subsection{Coarsening resistance and multicomponent solute transport}
\label{subsec:coarsening_resistance}

The coarsening of $\gamma^{\prime}$ precipitates was analysed within the framework of Ostwald ripening, where larger precipitates grow at the expense of smaller ones to reduce the total interfacial area. The earliest Lifshitz--Slyozov--Wagner (LSW) model was developed for dilute binary systems containing a vanishingly small precipitate volume fraction \cite{lifshitz1961kinetics,wagner1961theorie}. This dilute-limit assumption is not applicable to the present alloys, where the $\gamma^{\prime}$ volume fraction is greater than 0.6, ranging from $\phi=0.635$ for B--30Co to $\phi=0.753$ for B--10Co. In addition, the present Ni--Co--Cr--Al--Ti system is not a dilute binary alloy consisting of a single solute in a matrix, but a concentrated multicomponent $\gamma/\gamma^{\prime}$ alloy. Therefore, $\gamma^{\prime}$ coarsening involves the coupled redistribution of Ni, Cr, Co, Al, and Ti across the $\gamma/\gamma^{\prime}$ interface.

Subsequent LSW-based treatments account for finite precipitate volume fraction \cite{davies1980volume} and multicomponent diffusion effects \cite{philippe2013ostwald}. These modified theories generally retain the characteristic $t^{1/3}$ coarsening dependence, but the rate constant becomes more complex because it depends on precipitate volume fraction, interfacial energy, phase compositions, and the diffusivities of all solutes participating in redistribution across the $\gamma/\gamma^{\prime}$ interface. Therefore, the coarsening kinetics of the $\gamma^{\prime}$ precipitates in the present alloys were first evaluated from the temporal evolution of the mean precipitate radius using:

\begin{equation}
\langle r_t \rangle^3 - \langle r_0 \rangle^3 = K_{\mathrm{exp}} t
\label{eq:coarsening_cubic}
\end{equation}

where $\langle r_t \rangle$ is the mean $\gamma^{\prime}$ precipitate radius after ageing time $t$, $\langle r_0 \rangle$ is the initial mean precipitate radius, and $K_{\mathrm{exp}}$ is the experimentally measured coarsening-rate constant. For a concentrated multicomponent $\gamma/\gamma^{\prime}$ alloy, $K_{\mathrm{exp}}$ can be related to the dilute-limit LSW coefficient by including corrections for finite precipitate volume fraction and precipitate morphology:

\begin{equation}
K_{\mathrm{exp}} = A K_{\mathrm{LSW}} f(\phi)\chi
\label{eq:kexp_prefactor}
\end{equation}

where $A$ is the prefactor correction, $f(\phi)$ is the finite-volume-fraction correction factor, $\phi$ is the $\gamma^{\prime}$ volume fraction, and $\chi$ is the precipitate shape factor. The finite-volume-fraction correction factor was evaluated using the analytical correction adopted for high-volume-fraction $\gamma^{\prime}$ coarsening \cite{davies1980volume,yuan2025suppressing}. In this form, the volume-fraction-corrected coarsening coefficient is written as:

\begin{equation}
K_{\phi}=K_{\mathrm{LSW}}f(\phi)
\label{eq:k_phi}
\end{equation}

where $f(\phi)$ is given by:

\begin{equation}
f(\phi)
=
\frac{27}{4}
\left[
\frac{
2-\left(1-\sqrt{3\phi}\right)
\left(
1-\frac{1}{\sqrt{3\phi}}
+
\sqrt{
\frac{1}{3\phi}
+
\frac{1}{\sqrt{3\phi}}
+
1
}
\right)
}
{
\left(
1-\frac{1}{\sqrt{3\phi}}
+
\sqrt{
\frac{1}{3\phi}
+
\frac{1}{\sqrt{3\phi}}
+
1
}
\right)^3
}
\right].
\label{eq:f_phi}
\end{equation}

For the present alloys, Equation~\ref{eq:f_phi} gives $f(0.753)=3.431$ for B--10Co, $f(0.704)=3.348$ for B--20Co, and $f(0.635)=3.227$ for B--30Co. Thus, although the high $\gamma^{\prime}$ volume fraction must be explicitly accounted for, the correction varies only weakly across the present composition range.

The multicomponent LSW coefficient was expressed in terms of the resistance to solute redistribution through the $\gamma$ matrix:

\begin{equation}
K_{\mathrm{LSW}} =
\frac{8\sigma V_m}{9RT}
\left[
\sum_i
\frac{(\Delta C_i)^2}{D_i^{\gamma}C_i^{\gamma}}
\right]^{-1},
\qquad
i=\mathrm{Ni,Cr,Co,Al,Ti}
\label{eq:mlsw_transport}
\end{equation}

where $\sigma$ is the apparent $\gamma/\gamma^{\prime}$ interfacial-energy term, $V_m$ is the molar volume, $R$ is the universal gas constant, $T$ is the absolute temperature, $\Delta C_i = |C_i^{\gamma^{\prime}} - C_i^{\gamma}|$ is the composition difference of element $i$ between the $\gamma^{\prime}$ precipitate and the $\gamma$ matrix, $D_i^{\gamma}$ is the tracer diffusivity of element $i$ in the $\gamma$ matrix, and $C_i^{\gamma}$ is the corresponding matrix composition.

The summation term in Equation~\ref{eq:mlsw_transport} represents the cumulative resistance offered by all solutes that must redistribute during $\gamma^{\prime}$ coarsening. This multicomponent solute-transport resistance, $S$, is defined as:

\begin{equation}
S =
\sum_i
\frac{(\Delta C_i)^2}{D_i^{\gamma}C_i^{\gamma}}
\label{eq:S_definition}
\end{equation}

A high value of $S$ indicates that coarsening is kinetically restricted either because the relevant solute diffuses slowly in the $\gamma$ matrix, because it exhibits a large composition difference across the $\gamma/\gamma^{\prime}$ interface, or because both effects operate simultaneously. Rearranging Equation~\ref{eq:mlsw_transport} allows the apparent interfacial-energy term to be estimated from the experimentally measured coarsening rate:

\begin{equation}
\sigma =
K_{\mathrm{exp}}
\frac{9RT}{8V_m}
\frac{S}{A f(\phi)\chi}
\label{eq:sigma_compact}
\end{equation}

All calculations were carried out at \SI{900}{\celsius} using $T=\SI{1173.15}{\kelvin}$, $R=\SI{8.314}{\joule\per\mol\per\kelvin}$, and $V_m=\SI{7.12e-6}{\meter\cubed\per\mol}$. The prefactor correction, $A$, was evaluated from thermokinetic precipitation simulations performed using Thermo-Calc with the TCNI11 thermodynamic database and the MOBNI5 mobility database. In the simulations, the alloy compositions and ageing temperature were fixed, and simulated coarsening-rate constants were obtained for the corresponding $\gamma/\gamma^{\prime}$ systems. The simulated values were then compared iteratively with the experimentally measured coarsening-rate constants by adjusting the prefactor in Equation~\ref{eq:kexp_prefactor}. A common value of $A=1.67$ gave the best simultaneous agreement with the experimental coarsening rates of B--10Co and B--30Co and was therefore used for the apparent interfacial-energy evaluation.

\subsubsection{Solute-transport resistance and apparent interfacial-energy evaluation}

The calculated resistance terms show that Co addition changes the solute-transport pathway controlling $\gamma^{\prime}$ coarsening, as summarized in Table~\ref{tab:lsw_analysis}. In B--10Co, the total transport resistance is $8.811 \times 10^{15}~\mathrm{s\,m^{-2}}$, with the largest contributions arising from Cr and Ni. These elements contribute 35.4\% and 32.2\% of the total resistance, respectively, indicating that $\gamma^{\prime}$ coarsening in B--10Co is primarily controlled by redistribution within the Ni--Cr-rich $\gamma$ matrix. In contrast, the total resistance increases to $1.267 \times 10^{16}~\mathrm{s\,m^{-2}}$ in B--30Co, and Co becomes the largest contributor, accounting for 36.2\% of $S$. Ni and Cr still provide substantial contributions, but the dominant transport pathway shifts from Ni--Cr-controlled in B--10Co to Co-assisted Ni--Co--Cr-controlled in B--30Co.

\begin{table}[!htb]
\centering
\caption{Element-wise diffusivity and solute-transport resistance terms calculated for B--10Co and B--30Co aged at \SI{900}{\celsius}. Bold values indicate the largest resistance contributions to $S$. Detailed derivation and full resistance analysis are provided in Supplementary Note 1.}
\label{tab:lsw_analysis}
\scriptsize
\resizebox{\linewidth}{!}{
\begin{tabular}{llccccc}
\toprule
\textbf{Alloy} &
\textbf{Element} &
\textbf{$D_i^\gamma$} &
\textbf{$\Delta C_i$} &
\textbf{$D_i^\gamma C_i^\gamma$} &
\textbf{$S_i=(\Delta C_i)^2/(D_i^\gamma C_i^\gamma)$} &
\textbf{Contribution to $S$} \\
&
&
\textbf{(\si{\meter\squared\per\second})} &
&
\textbf{(\si{\meter\squared\per\second})} &
\textbf{(\si{\second\per\meter\squared})} &
\textbf{(\%)} \\
\midrule
\textbf{B--10Co} & Ni & $\mathbf{5.62 \times 10^{-17}}$ & 0.253 & $2.259 \times 10^{-17}$ & $\mathbf{2.834 \times 10^{15}}$ & \textbf{32.2} \\
& Cr & $\mathbf{8.35 \times 10^{-17}}$ & 0.285 & $2.605 \times 10^{-17}$ & $\mathbf{3.118 \times 10^{15}}$ & \textbf{35.4} \\
& Co & $1.01 \times 10^{-16}$ & 0.142 & $2.101 \times 10^{-17}$ & $9.597 \times 10^{14}$ & 10.9 \\
& Al & $2.79 \times 10^{-16}$ & 0.105 & $1.060 \times 10^{-17}$ & $1.040 \times 10^{15}$ & 11.8 \\
& Ti & $2.44 \times 10^{-16}$ & 0.078 & $0.7076 \times 10^{-17}$ & $8.598 \times 10^{14}$ & 9.8 \\
\cmidrule{2-7}
& \textbf{Total} & -- & -- & -- & $\mathbf{8.811 \times 10^{15}}$ & \textbf{100} \\
\midrule
\textbf{B--30Co} & Ni & $\mathbf{6.38 \times 10^{-17}}$ & 0.231 & $1.461 \times 10^{-17}$ & $\mathbf{3.652 \times 10^{15}}$ & \textbf{28.8} \\
& Cr & $7.29 \times 10^{-17}$ & 0.217 & $1.874 \times 10^{-17}$ & $2.513 \times 10^{15}$ & 19.8 \\
& Co & $\mathbf{1.58 \times 10^{-17}}$ & 0.179 & $0.6984 \times 10^{-17}$ & $\mathbf{4.588 \times 10^{15}}$ & \textbf{36.2} \\
& Al & $4.37 \times 10^{-16}$ & 0.070 & $2.229 \times 10^{-17}$ & $2.198 \times 10^{14}$ & 1.7 \\
& Ti & $2.37 \times 10^{-16}$ & 0.094 & $0.5214 \times 10^{-17}$ & $1.695 \times 10^{15}$ & 13.4 \\
\cmidrule{2-7}
& \textbf{Total} & -- & -- & -- & $\mathbf{1.267 \times 10^{16}}$ & \textbf{100} \\
\bottomrule
\end{tabular}
}
\end{table}

This change is important because Co is not merely a passive matrix-forming element in the present alloys. Although Co partitions less strongly to $\gamma^{\prime}$ than Ti or Al, the absolute composition difference of Co across the $\gamma/\gamma^{\prime}$ interface becomes large in B--30Co. Combined with its reduced diffusivity in the Co-rich $\gamma$ matrix, this makes Co a major contributor to the solute flux required for precipitate coarsening. Therefore, the improved coarsening resistance of B--30Co arises from a collective multicomponent transport effect rather than from a single slow-diffusing solute.

Using the same correction factors, the apparent $\gamma/\gamma^{\prime}$ interfacial-energy term decreases from \SI{25.0}{\milli\joule\per\meter\squared} for B--10Co to \SI{7.55}{\milli\joule\per\meter\squared} for B--30Co, as shown in Table~\ref{tab:sigma_final}. The finite-volume-fraction correction varies only slightly over this composition range, decreasing from $f(\phi)=3.431$ for B--10Co to $f(\phi)=3.227$ for B--30Co. The ratio $f(0.753)/f(0.635)=1.063$ indicates that the difference in $\gamma^{\prime}$ volume fraction would account for only a small change in the coarsening rate. This is much smaller than the experimentally observed coarsening-rate ratio, $K_r^{\mathrm{B\text{--}10Co}}/K_r^{\mathrm{B\text{--}30Co}} = 8.98/1.73 = 5.19$.
Thus, the large reduction in coarsening rate with increasing Co content cannot be explained by finite-volume-fraction effects alone. Instead, it results from the simultaneous increase in multicomponent solute-transport resistance and decrease in the apparent interfacial-energy term.

\begin{table}[!htb]
\centering
\caption{Prefactor-corrected apparent $\gamma/\gamma^{\prime}$ interfacial-energy term calculated for alloys aged at \SI{900}{\celsius}.}
\label{tab:sigma_final}
\resizebox{0.7\linewidth}{!}{%
\begin{tabular}{lcccccc}
\toprule
\textbf{Alloy} &
\textbf{$K_{\mathrm{exp}}$} &
\textbf{$S$} &
\textbf{$\phi$} &
\textbf{$f(\phi)$} &
\textbf{$\chi$} &
\textbf{$\sigma$} \\
&
\textbf{(\si{\nano\meter\cubed\per\second})} &
\textbf{(\si{\second\per\meter\squared})} &
&
&
&
\textbf{(\si{\milli\joule\per\meter\squared})} \\
\midrule
\textbf{B--10Co} & 8.98 & $8.811 \times 10^{15}$ & 0.753 & 3.431 & 0.85 & 25.0 \\
\textbf{B--20Co} & 4.28 & $8.529 \times 10^{15}$ & 0.704 & 3.348 & 0.84 & 13.2 \\
\textbf{B--30Co} & 1.73 & $1.267 \times 10^{16}$ & 0.635 & 3.227 & 0.83 & 7.55 \\
\bottomrule
\end{tabular}%
}
\end{table}

The $\sigma$ values obtained here should therefore be treated as apparent interfacial-energy terms rather than equilibrium $\gamma/\gamma^{\prime}$ interfacial energies, because the calculation uses tracer diffusivities and assumes diagonal transport without explicitly considering cross-diffusion or thermodynamic interaction terms. Nevertheless, the analysis provides a useful comparative measure of how Co addition modifies the kinetic pathway for $\gamma^{\prime}$ coarsening. Overall, increasing Co content transforms the coarsening process from a primarily Ni--Cr-controlled transport pathway in B--10Co to a Co-assisted multicomponent transport-controlled pathway in B--30Co, thereby stabilizing the $\gamma/\gamma^{\prime}$ microstructure at \SI{900}{\celsius}.

\section{Conclusions}

This work examined the effect of increasing Co content in Ti-rich Ni--Co--Cr--Al--Ti $\gamma/\gamma^\prime$ superalloys on the $\gamma^\prime$ coarsening behaviour. The following conclusions can be drawn from the present work:

\begin{itemize}

    \item All three alloys show a two-phase $\gamma/\gamma^\prime$ microstructure after heat treatment, without detectable secondary phases within the resolution of the present SEM/XRD analysis. Increasing Co from B--10Co to B--30Co refines the mean $\gamma^\prime$ precipitate radius from $\sim\SI{106}{\nano\meter}$ to $\sim\SI{75}{\nano\meter}$, while reducing the $\gamma^\prime$ volume fraction from $\sim\SI{75.3}{\percent}$ to $\sim\SI{63.5}{\percent}$. APT analysis shows that the $\gamma^\prime$ precipitates retain an A$_3$B-type L1$_2$ stoichiometry close to $\text{(Ni,Co,Cr)}_3\text{(Al,Ti)}$. With increasing Co, Ti partitioning to $\gamma^\prime$ increases, whereas Al partitioning decreases, indicating progressive Ti substitution on the B sublattice of the L1$_2$ unit cell. Concurrently, the $\gamma$ matrix evolves from a relatively Ni--Cr-rich matrix in B--10Co, with an approximate Ni:Co:Cr ratio of 2:1:2, to a more balanced Ni--Co--Cr matrix in B--20Co, with Ni:Co:Cr $\sim$2:2:1, and finally to a Co-enriched matrix in B--30Co, with Ni:Co:Cr $\sim$1:2:1. Thus, systematic Co-for-Ni substitution establishes distinct MEA-type $\gamma$-matrix environments across the alloy series.

    \item Increasing Co content decreases the $\gamma^\prime$ solvus temperature from $\sim\SI{1250}{\celsius}$ in B--10Co to $\sim\SI{1185}{\celsius}$ in B--30Co. At the same time, the constrained $\gamma/\gamma^\prime$ lattice misfit increases slightly from approximately $+0.25\%$ to $+0.28\%$. Thus, B--30Co exhibits better coarsening resistance despite having a lower $\gamma^\prime$ solvus temperature and a slightly higher positive lattice misfit, showing that solvus temperature and lattice misfit alone do not control the coarsening response.

    \item B--30Co exhibits the slowest $\gamma^\prime$ coarsening kinetics among the investigated alloys, with coarsening rate constants of $\sim\SI{1.73}{\nano\meter\cubed\per\second}$ at \SI{900}{\celsius}, $\sim\SI{3.85}{\nano\meter\cubed\per\second}$ at \SI{950}{\celsius}, and $\sim\SI{7.46}{\nano\meter\cubed\per\second}$ at \SI{1000}{\celsius}. The apparent activation energy for coarsening increases from $\sim\SI{155.7}{\kilo\joule\per\mol}$ for B--10Co to $\sim\SI{302.0}{\kilo\joule\per\mol}$ for B--30Co, confirming that Co enrichment in the $\gamma$ matrix strongly modifies the solute-redistribution pathway during $\gamma^\prime$ coarsening.
\item The prefactor-corrected multicomponent LSW analysis shows that coarsening resistance is governed by the combined effects of elemental diffusivity and $\gamma/\gamma^\prime$ compositional partitioning through the transport-resistance term, \(S=\sum_i\frac{(\Delta C_i)^2}{D_i^{\gamma}C_i^{\gamma}}\). The total transport resistance increases from $\sim 8.811\times10^{15}$~s~m$^{-2}$ for B--10Co to $\sim 1.267\times10^{16}$~s~m$^{-2}$ for B--30Co, with the dominant resistance pathway shifting from primarily Ni--Cr controlled in B--10Co to Ni--Co--Cr controlled in B--30Co. Within the same framework, the apparent $\gamma/\gamma^\prime$ interfacial energy decreases from $\sim\SI{25.0}{\milli\joule\per\meter\squared}$ to $\sim\SI{7.55}{\milli\joule\per\meter\squared}$, indicating a reduced apparent capillarity contribution in the Co-rich alloy.

\end{itemize}

\section*{Data and code availability}
Relevant Codes can be accessed at : https://l1nq.com/pqjjb86 (SM's Github repository), it also includes the relevant simulation or plotting data as applicable. 

\section*{Acknowledgments}
S.M. thanks Dr. Hemant Kumar, Department of Materials Engineering, IISc, and Ms. Archana S., AFMM, IISc, for their assistance with FIB/APT training and the initial APT experiments .

\section*{Funding Statement}
The authors acknowledge the Advanced Facility for Microscopy and Microanalysis (AFMM), Indian Institute of Science (IISc), Bangalore, funded by the DST-FIST program, Government of India, for access to microscopy and tomography facilities used in the work. \textbf{S.K.M. and S.M.} acknowledge financial support from the MPG–IISc Partner Group. \textbf{S.M.} acknowledges the DST Prime Minister’s Research Fellowship (PMRF).

\section*{Author Contributions}

\begin{itemize}
    \item \textbf{S.M.:} Conceptualization; methodology; alloy design / preparation; microscopy, microanalysis; data analysis and interpretation; writing--original draft.
    
    \item \textbf{S.S.:} resources; supervision; writing--review and editing.
    
    \item \textbf{B.S.M.:} Interpretation of thermodynamic modelling; computational resources; supervision; writing--review and editing.
    
    \item \textbf{S.K.M.:} Validation and interpretation of atom-probe-tomography data and superalloy physical metallurgy; resources; supervision; writing--review and editing.
\end{itemize}

All authors reviewed and approved the final manuscript.

\section*{Declaration of Competing Interest}

The authors declare no competing interests. The manuscript is original, has been approved by all authors, and is not under consideration elsewhere.

% =========================================================
% SUPPLEMENTARY MATERIAL
% =========================================================

\clearpage
\appendix

\begin{center}

{\LARGE\bfseries Supplementary Material\par}

\vspace{1.2em}

{\Large\bfseries
Cobalt-Controlled Interphase Partitioning Regulates Matrix Solute Transport\\
and $\gamma^\prime$ Coarsening in Ti-Rich NiCoCr-Based Superalloys
\par}

\vspace{1.5em}

{\large
Sudeepta Mukherjee$^{1,*}$,
Surendra Kumar Makineni$^{1}$,
B.~S. Murty$^{2}$,
Satyam Suwas$^{1,*}$
\par}

\vspace{1em}

{\normalsize
$^{1}$Department of Materials Engineering, Indian Institute of Science,
Bengaluru, Karnataka 560012, India
\par}

\vspace{0.4em}

{\normalsize
$^{2}$Department of Materials Science and Metallurgical Engineering,
Indian Institute of Technology Hyderabad, Kandi, Sangareddy,
Telangana 502285, India
\par}

\vspace{1em}

{\normalsize
$^{*}$Corresponding authors:
Sudeepta Mukherjee
(\href{mailto:sudeeptam@iisc.ac.in}{sudeeptam@iisc.ac.in});
Satyam Suwas
(\href{mailto:satyamsuwas@iisc.ac.in}{satyamsuwas@iisc.ac.in})
\par}

\end{center}

\vspace{2em}

% =========================================================
% SUPPLEMENTARY NUMBERING
% =========================================================

\setcounter{figure}{0}
\setcounter{table}{0}
\setcounter{equation}{0}

% Produces Figure S1, Table S1, and Equation S1.
\renewcommand{\thefigure}{S\arabic{figure}}
\renewcommand{\thetable}{S\arabic{table}}
\renewcommand{\theequation}{S\arabic{equation}}

% Table captions use the normal Roman document font.
\captionsetup[table]{
    labelfont=bf,
    justification=justified,
    singlelinecheck=false,
    skip=\medskipamount
}

% Manual italic caption for multi-page supplementary figures.
% This does not invoke \caption or \caption*, thereby avoiding
% duplicate labels such as "Figure Figure S1".
%
% Usage:
% \suppcaption{(i)}{Caption text}
%
% Output:
% Figure S1(i): Caption text
\newcommand{\suppcaption}[2]{%
    \par\medskip
    \begingroup
        \normalfont
        \rmfamily
        \itshape
        \normalsize
        \singlespacing
        \justifying
        \noindent
        \textbf{\figurename~\thefigure#1:}~#2\par
    \endgroup
}

% =========================================================
% TABLE S1
% Nominal and measured alloy compositions
% =========================================================

\begin{table}[!htb]
\centering
\normalfont
\rmfamily

\caption{Nominal and measured compositions obtained by WDS for the B--10Co, B--20Co, and B--30Co alloys and their corresponding designations.}
\label{tab:S1_wds}

\renewcommand{\arraystretch}{1.35}

\resizebox{\linewidth}{!}{%
\begin{tabular}{l l l}
\toprule

\textbf{Nominal composition (at.\%)} &
\textbf{Alloy designation} &
\textbf{Measured composition (at.\%)} \\

\midrule

\multirow{3}{*}{%
    Ni--11Cr--$x$Co--9Al--8Ti, 
    $x=10$, 20, and 30
}
&
B--10Co
&
Ni--bal.\ Cr--10.9 $\pm$ 0.3 Co--9.1 $\pm$ 0.2
Al--8.2 $\pm$ 0.1 Ti--8.0 $\pm$ 0.2
\\

&
B--20Co
&
Ni--bal.\ Cr--11.0 $\pm$ 0.2 Co--19.8 $\pm$ 0.3
Al--9.0 $\pm$ 0.2 Ti--8.1 $\pm$ 0.1
\\

&
B--30Co
&
Ni--bal.\ Cr--11.3 $\pm$ 0.2 Co--29.2 $\pm$ 0.4
Al--8.9 $\pm$ 0.2 Ti--7.9 $\pm$ 0.1
\\

\bottomrule
\end{tabular}%
}

\end{table}

% =========================================================
% FIGURE S1
% Temporal SEM micrographs
% =========================================================

% Increment the figure counter only once for all parts of Figure S1.
\refstepcounter{figure}
\phantomsection
\label{fig:S1_temporal_sem}

% ------------------------- Figure S1(i) -------------------

\begin{figure}[!htbp]
    \centering

    \includegraphics[
        width=\linewidth,
        height=0.82\textheight,
        keepaspectratio
    ]{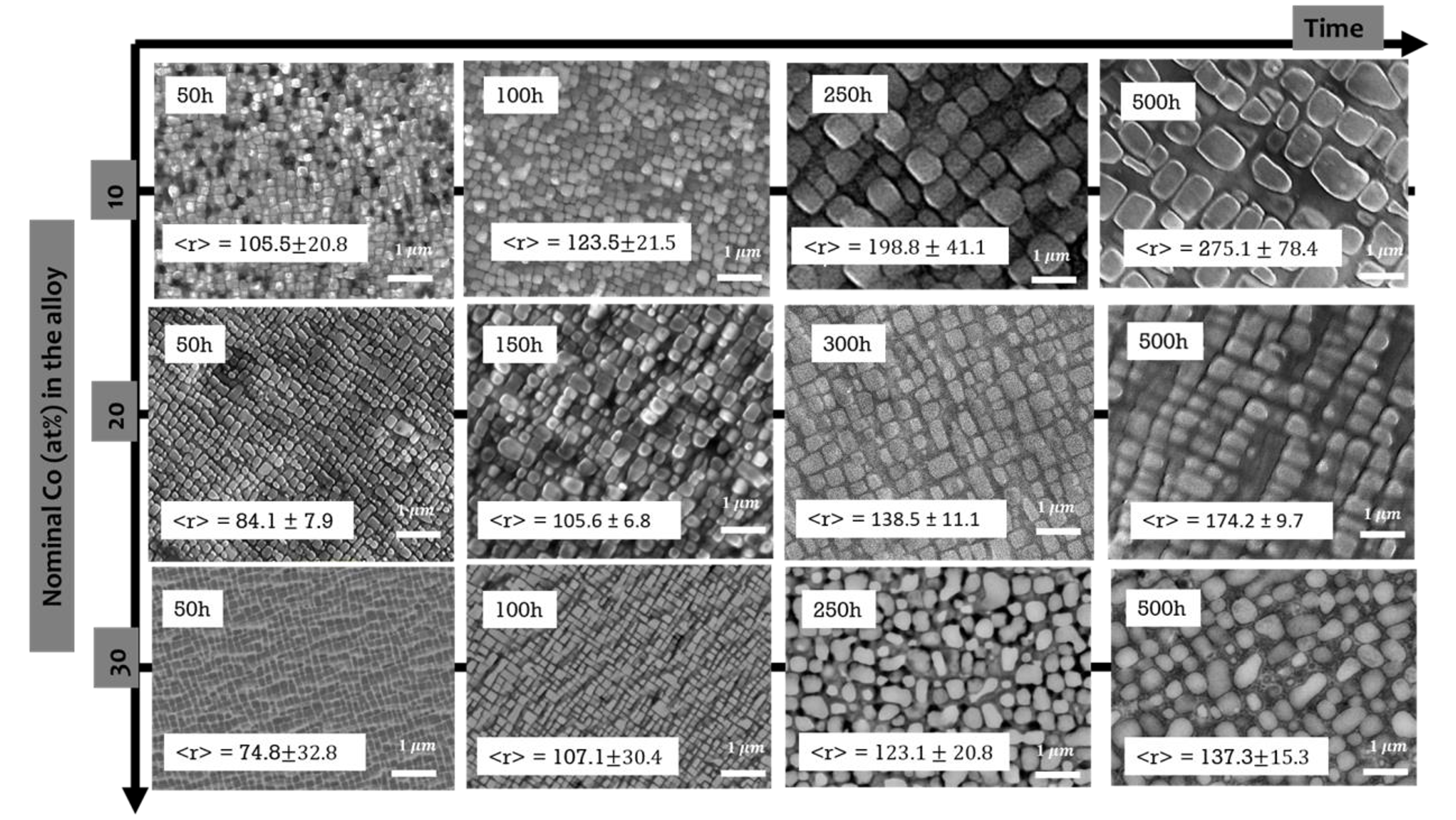}

    \suppcaption{(i)}
    {SEM micrographs of B--10Co, B--20Co, and B--30Co alloys isothermally heat-treated at \SI{900}{\degreeCelsius}, showing their microstructural evolution during ageing for 50, 100, 250, and 500~h. All micrographs were acquired at a magnification of 50\,000$\times$ to permit direct comparison.}

    \phantomsection
    \label{fig:S1_i}
\end{figure}

% ------------------------- Figure S1(ii) ------------------

\begin{figure}[!htbp]
    \centering

    \includegraphics[
        width=\linewidth,
        height=0.82\textheight,
        keepaspectratio
    ]{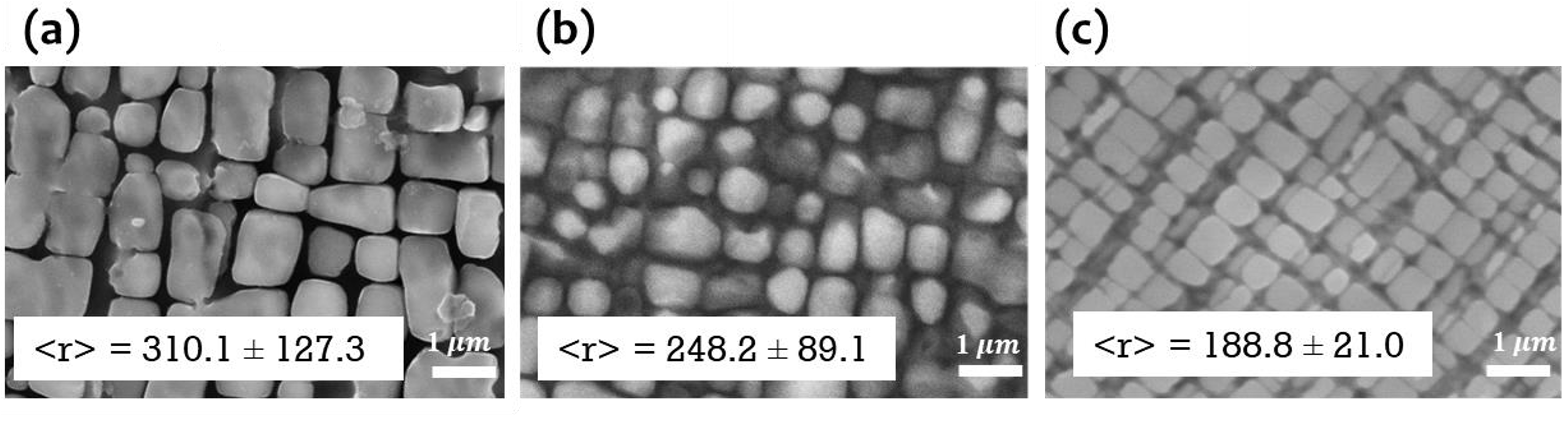}

    \suppcaption{(ii)}
    {SEM micrographs of the alloys after ageing at \SI{900}{\degreeCelsius} for 1000~h: (a) B--10Co, (b) B--20Co, and (c) B--30Co.}

    \phantomsection
    \label{fig:S1_ii}
\end{figure}

% ------------------------- Figure S1(iii) -----------------

\begin{figure}[!htbp]
    \centering

    \includegraphics[
        width=\linewidth,
        height=0.82\textheight,
        keepaspectratio
    ]{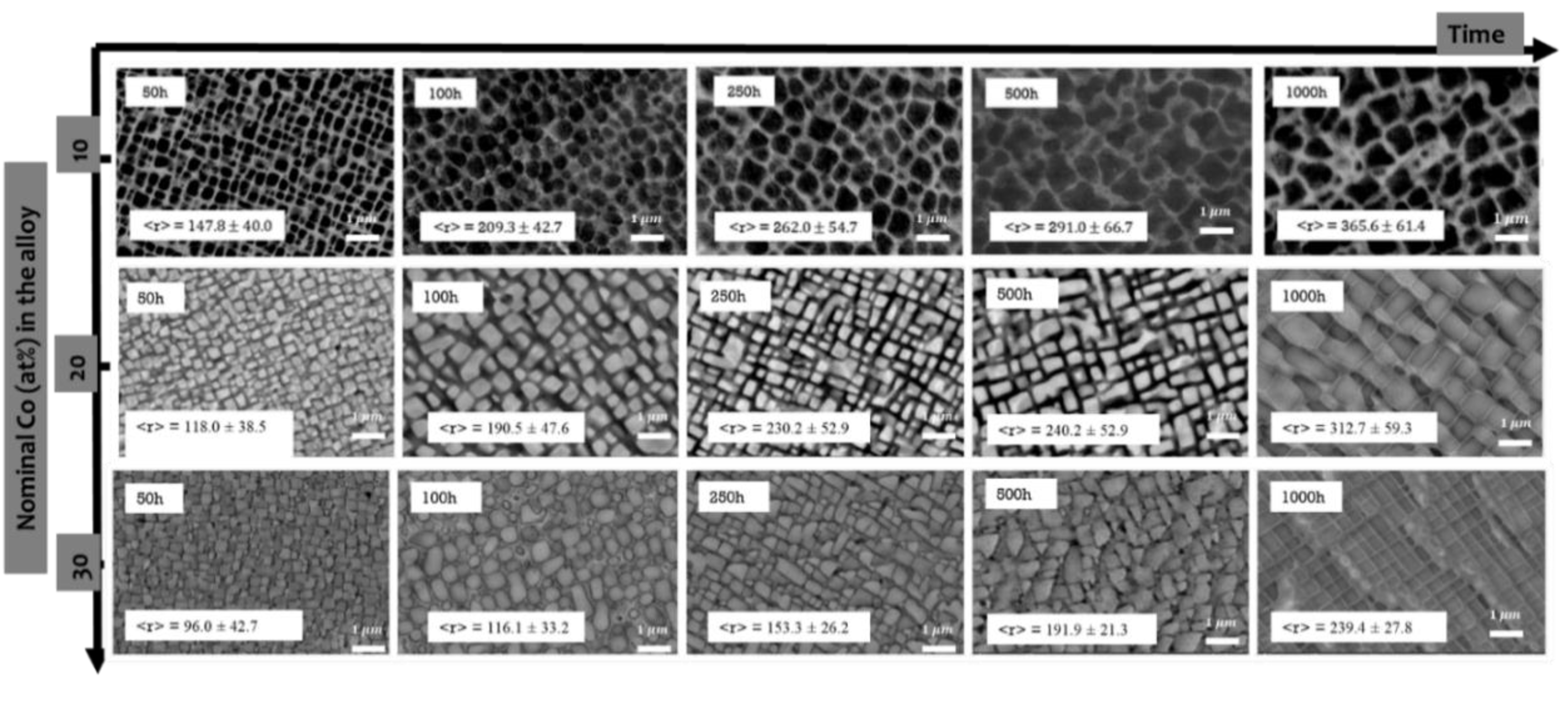}

    \suppcaption{(iii)}
    {SEM micrographs of B--10Co, B--20Co, and B--30Co alloys isothermally heat-treated at \SI{950}{\degreeCelsius}, showing their microstructural evolution during ageing for 50, 100, 250, 500, and 1000~h.}

    \phantomsection
    \label{fig:S1_iii}
\end{figure}

% ------------------------- Figure S1(iv) ------------------

\begin{figure}[!htbp]
    \centering

    \includegraphics[
        width=\linewidth,
        height=0.82\textheight,
        keepaspectratio
    ]{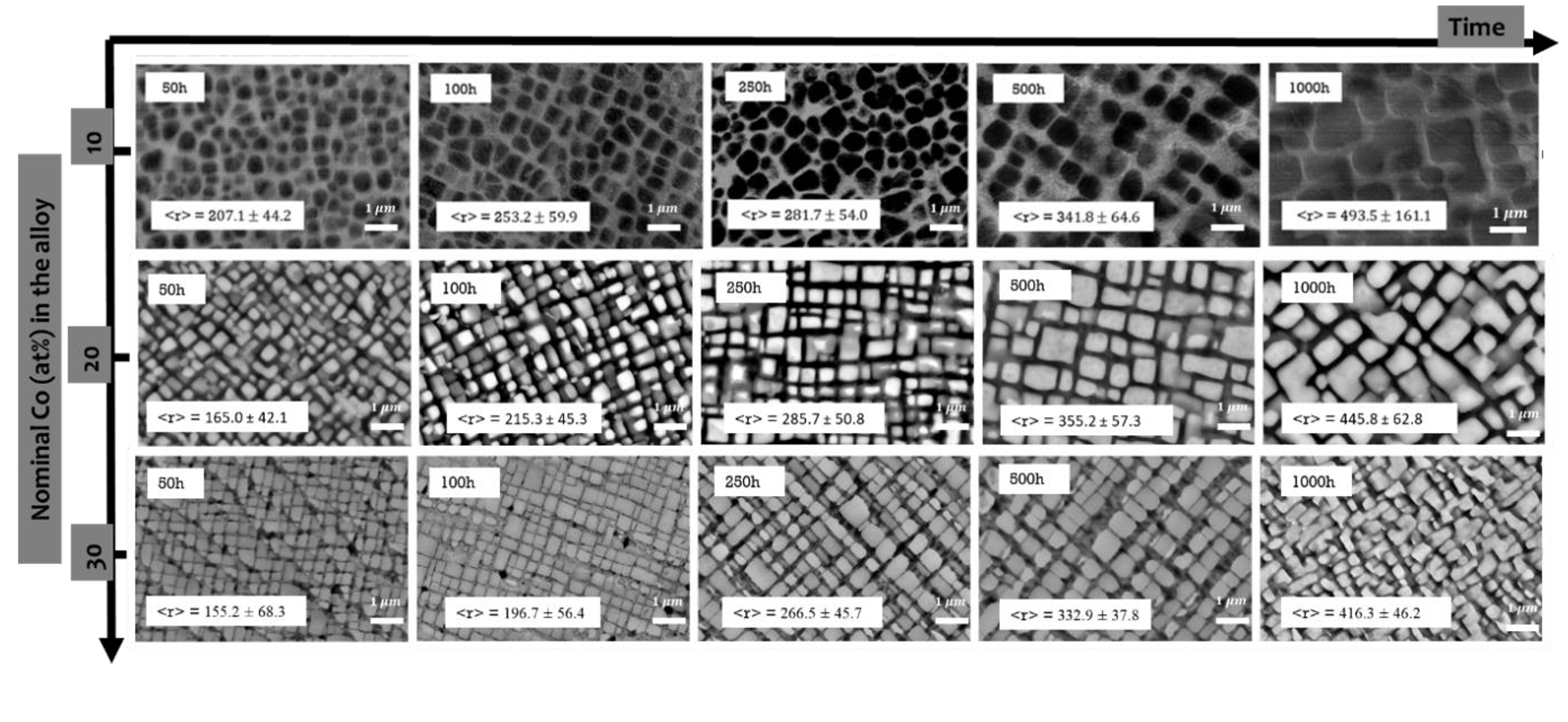}

    \suppcaption{(iv)}
    {SEM micrographs of B--10Co, B--20Co, and B--30Co alloys isothermally heat-treated at \SI{1000}{\degreeCelsius}, showing their microstructural evolution during ageing for 50, 100, 250, 500, and 1000~h.}

    \phantomsection
    \label{fig:S1_iv}
\end{figure}

% =========================================================
% FIGURE S2
% Temporal evolution of average precipitate size
% =========================================================

% Increment the figure counter only once for all parts of Figure S2.
\refstepcounter{figure}
\phantomsection
\label{fig:S2_temporal_evolution}

% ------------------------- Figure S2(i) -------------------

\begin{figure}[!htbp]
    \centering

    \includegraphics[
        width=\linewidth,
        height=0.42\textheight,
        keepaspectratio
    ]{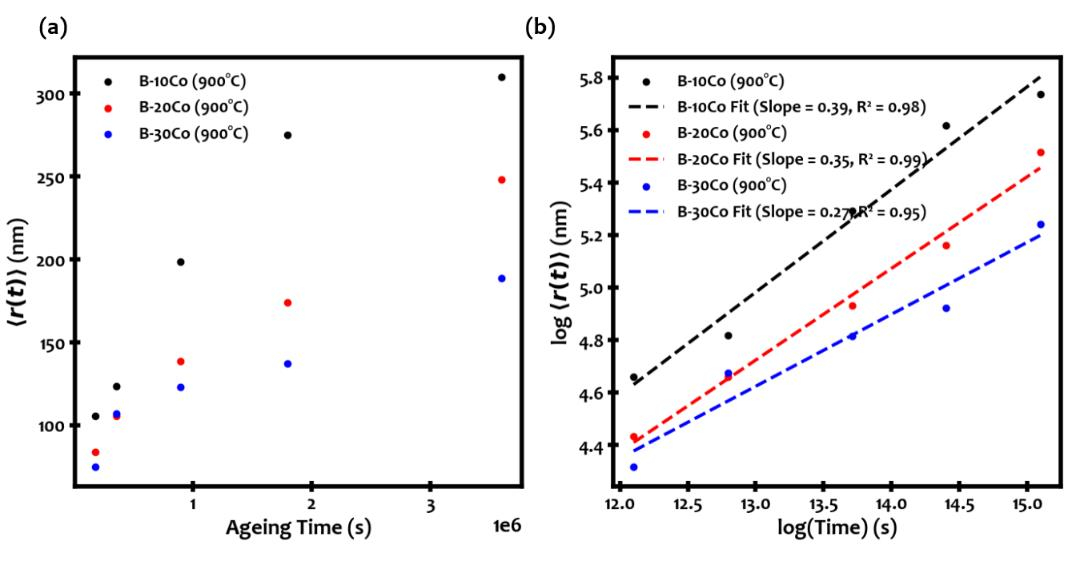}

    \par\medskip

    \includegraphics[
        width=\linewidth,
        height=0.42\textheight,
        keepaspectratio
    ]{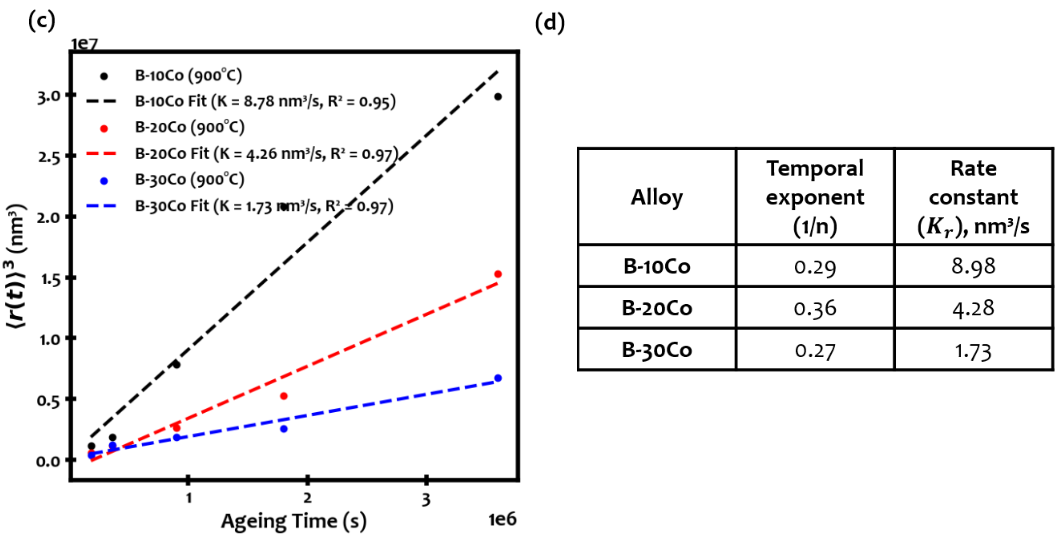}

    \suppcaption{(i)}
    {Temporal evolution of the $\gamma^\prime$ precipitates in the B--10Co, B--20Co, and B--30Co alloys during ageing at \SI{900}{\degreeCelsius}. The plots show: (a) the area-equivalent precipitate radius, $\langle r(t)\rangle$, as a function of ageing time; (b) $\log\langle r(t)\rangle$ versus $\log t$, showing the temporal exponent of the $\gamma^\prime$ precipitate-size evolution relative to the classical and modified LSW models; and (c) $\langle r(t)\rangle^{3}$ as a function of ageing time, with linear regression fits used to determine the coarsening-rate constants, $K_r$.}

    \phantomsection
    \label{fig:S2_i}
\end{figure}

% ------------------------- Figure S2(ii) ------------------

\begin{figure}[!htbp]
    \centering

    \includegraphics[
        width=\linewidth,
        height=0.42\textheight,
        keepaspectratio
    ]{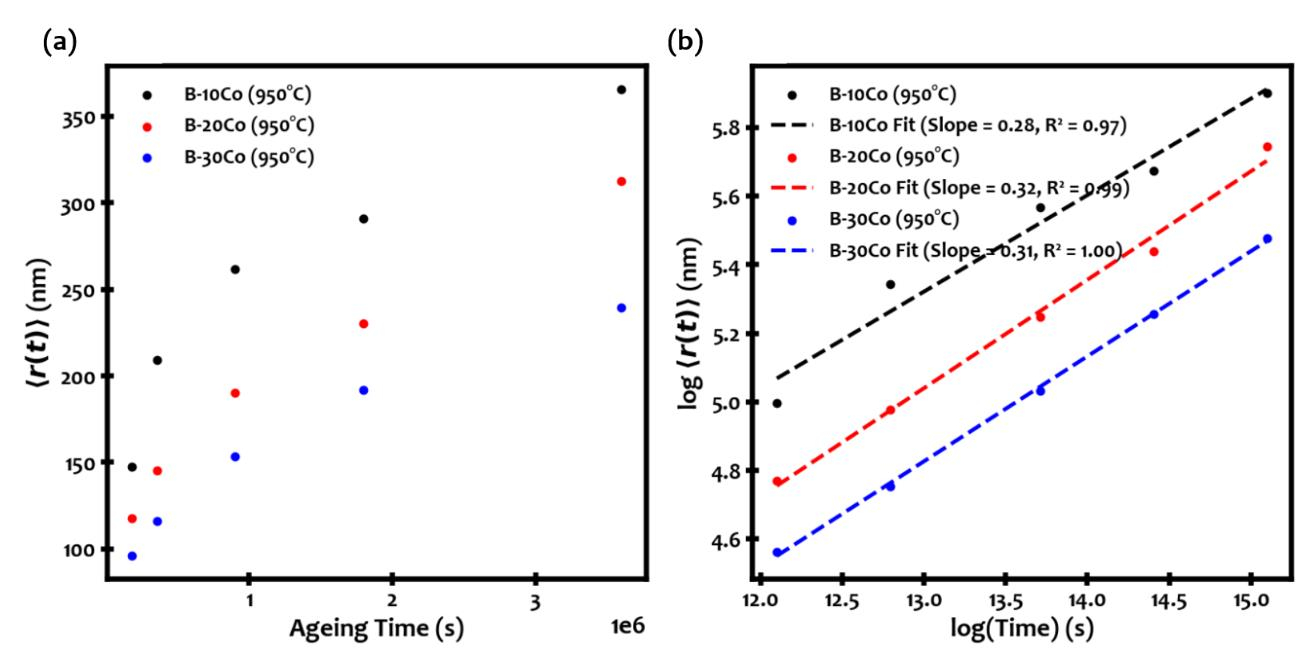}

    \par\medskip

    \includegraphics[
        width=\linewidth,
        height=0.42\textheight,
        keepaspectratio
    ]{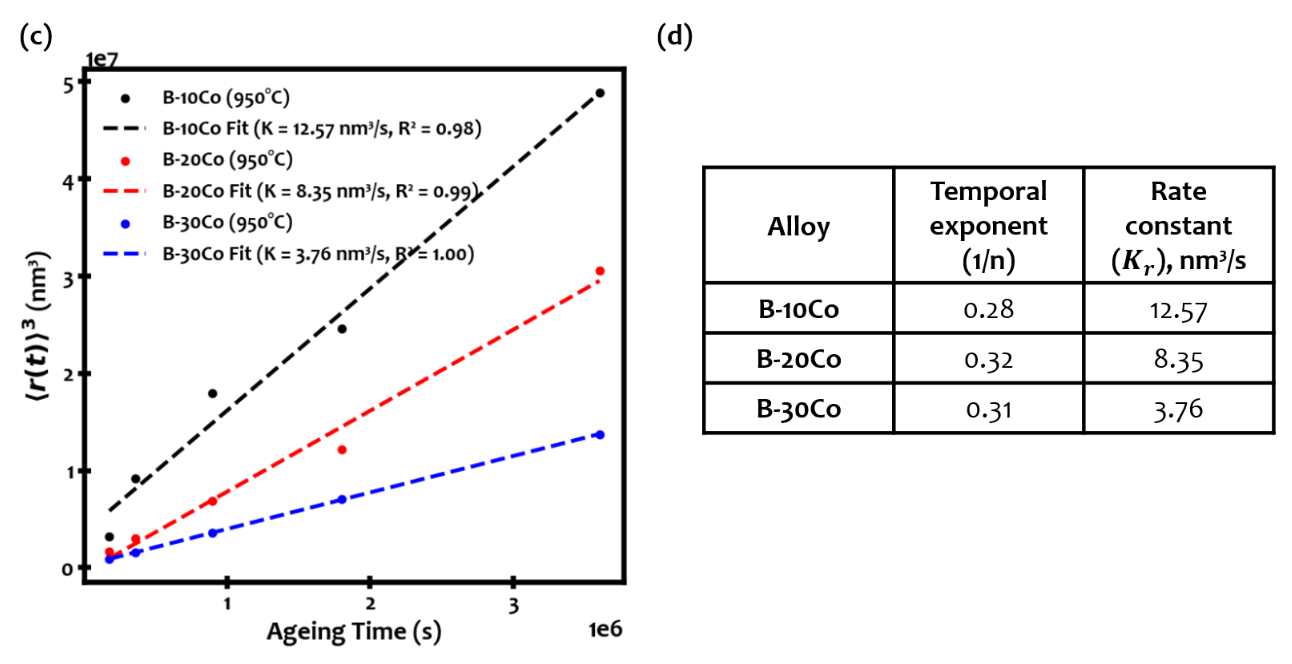}

    \suppcaption{(ii)}
    {Temporal evolution of the $\gamma^\prime$ precipitates in the B--10Co, B--20Co, and B--30Co alloys during ageing at \SI{950}{\degreeCelsius}. The plots show: (a) the area-equivalent precipitate radius, $\langle r(t)\rangle$, as a function of ageing time; (b) $\log\langle r(t)\rangle$ versus $\log t$, showing the temporal exponent of the $\gamma^\prime$ precipitate-size evolution relative to the classical and modified LSW models; and (c) $\langle r(t)\rangle^{3}$ as a function of ageing time, with linear regression fits used to determine the coarsening-rate constants, $K_r$.}

    \phantomsection
    \label{fig:S2_ii}
\end{figure}

% ------------------------- Figure S2(iii) -----------------

\begin{figure}[!htbp]
    \centering

    \includegraphics[
        width=\linewidth,
        height=0.42\textheight,
        keepaspectratio
    ]{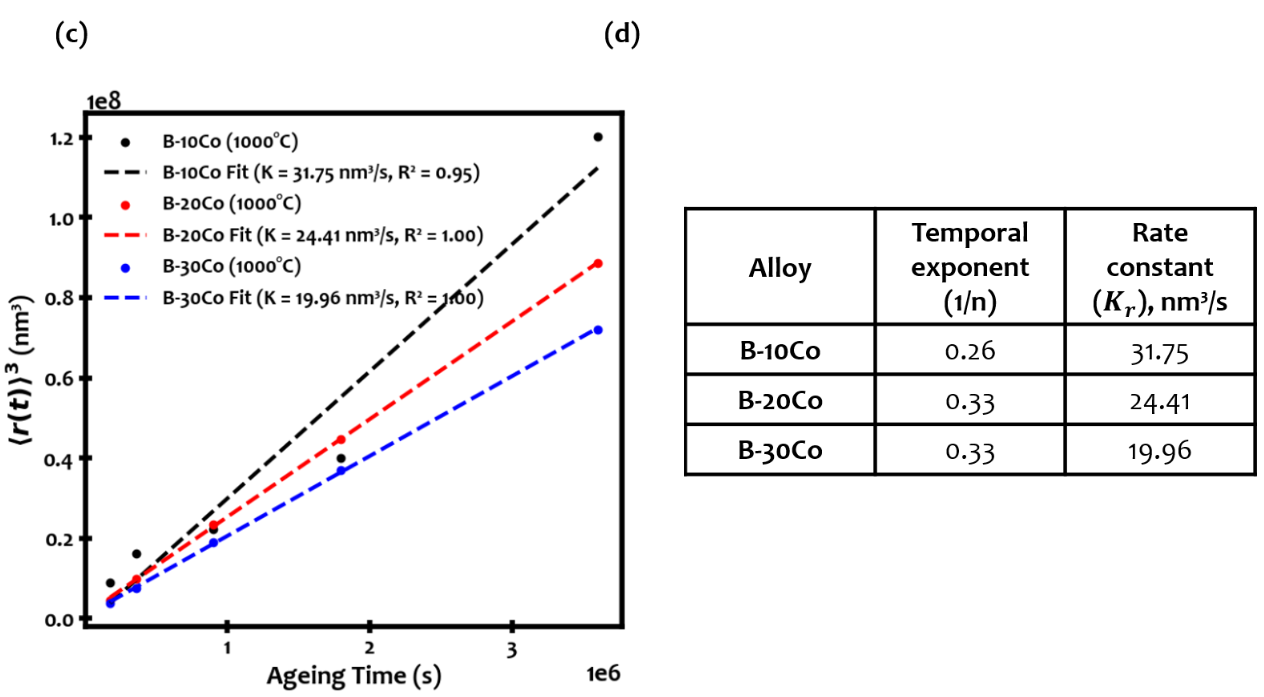}

    \par\medskip

    \includegraphics[
        width=\linewidth,
        height=0.42\textheight,
        keepaspectratio
    ]{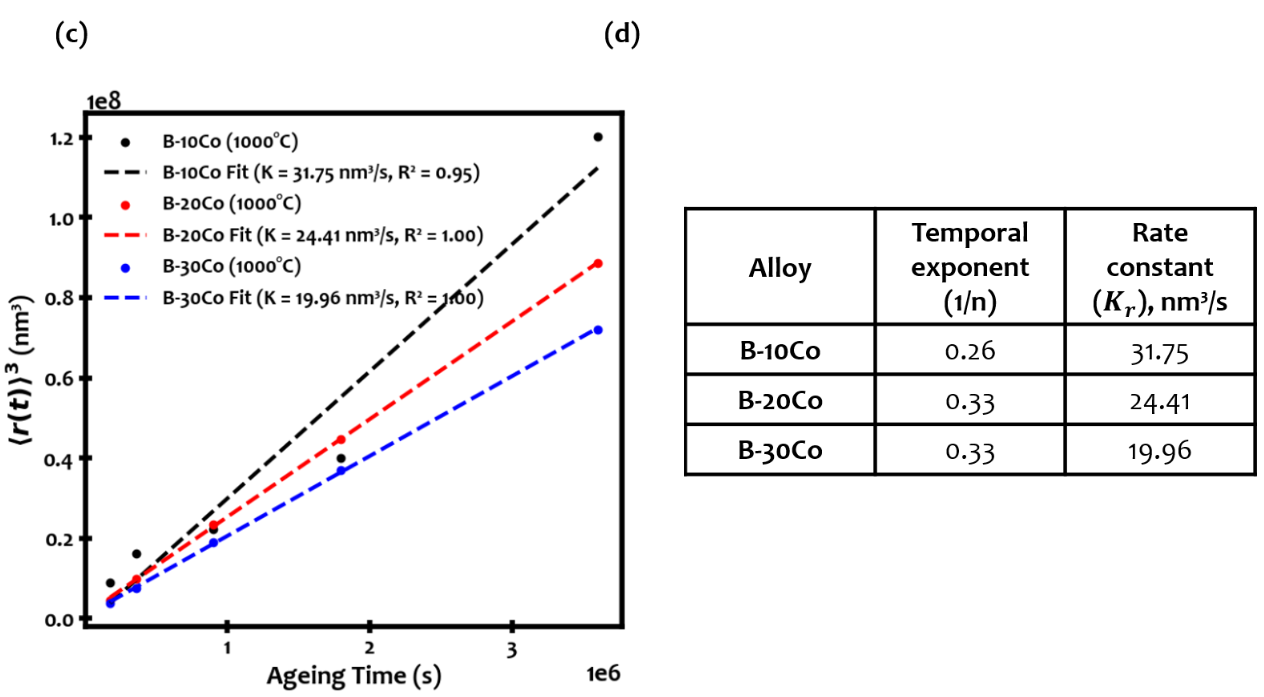}

    \suppcaption{(iii)}
    {Temporal evolution of the $\gamma^\prime$ precipitates in the B--10Co, B--20Co, and B--30Co alloys during ageing at \SI{1000}{\degreeCelsius}. The plots show: (a) the area-equivalent precipitate radius, $\langle r(t)\rangle$, as a function of ageing time; (b) $\log\langle r(t)\rangle$ versus $\log t$, showing the temporal exponent of the $\gamma^\prime$ precipitate-size evolution relative to the classical and modified LSW models; and (c) $\langle r(t)\rangle^{3}$ as a function of ageing time, with linear regression fits used to determine the coarsening-rate constants, $K_r$.}

    \phantomsection
    \label{fig:S2_iii}
\end{figure}

% =========================================================
% FIGURE S3
% Precipitate-size-distribution evolution
% =========================================================

\begin{landscape}

% Increment the figure counter only once for all parts of Figure S3.
\refstepcounter{figure}
\phantomsection
\label{fig:S3_psd_evolution}

% ------------------------- Figure S3(i) -------------------

\begin{figure}[!htbp]
    \centering

    \includegraphics[
        width=\linewidth,
        height=0.82\textheight,
        keepaspectratio
    ]{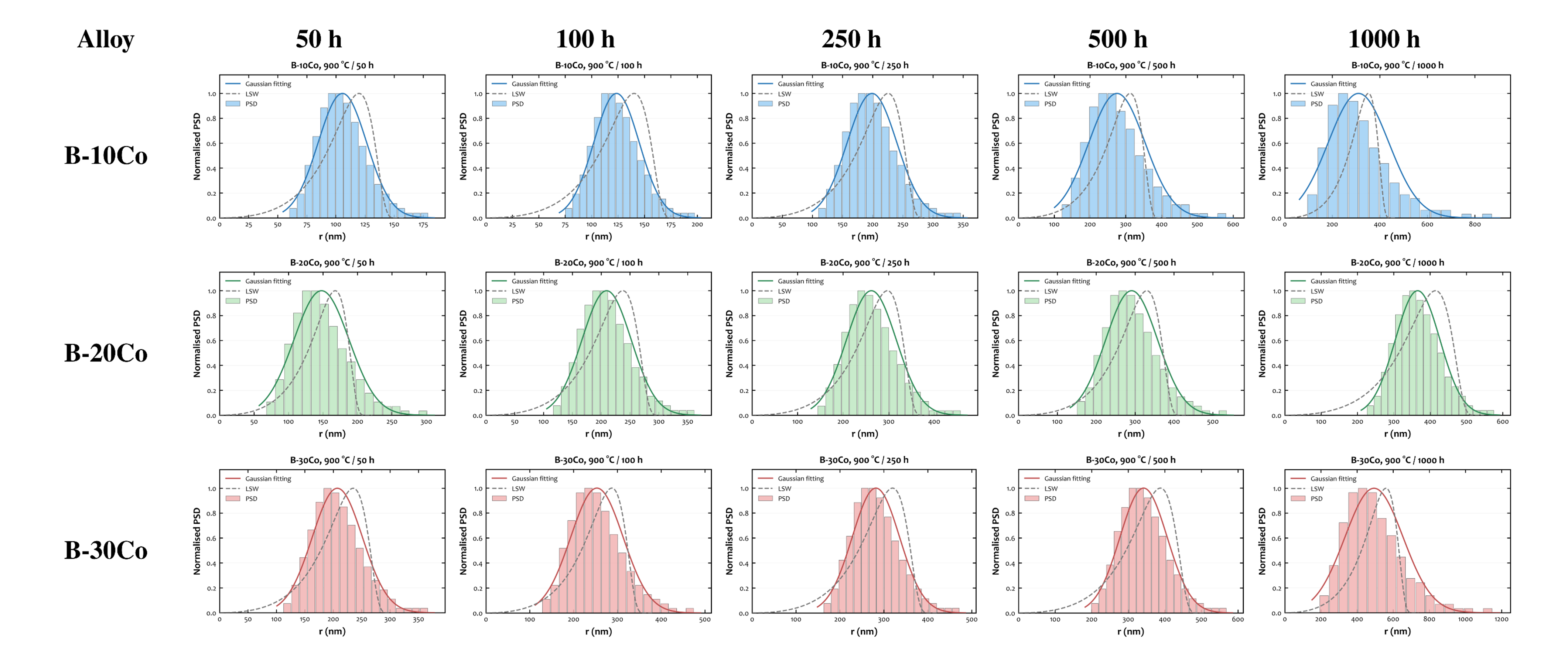}

    \suppcaption{(i)}
    {Precipitate-size-distribution evolution of the B--10Co, B--20Co, and B--30Co alloys aged at \SI{900}{\degreeCelsius} for 50, 100, 250, 500, and 1000~h. Each plot shows the normalized PSD as a function of precipitate radius, together with the corresponding Gaussian fit and LSW distribution.}

    \phantomsection
    \label{fig:S3_i}
\end{figure}

% ------------------------- Figure S3(ii) ------------------

\begin{figure}[!htbp]
    \centering

    \includegraphics[
        width=\linewidth,
        height=0.82\textheight,
        keepaspectratio
    ]{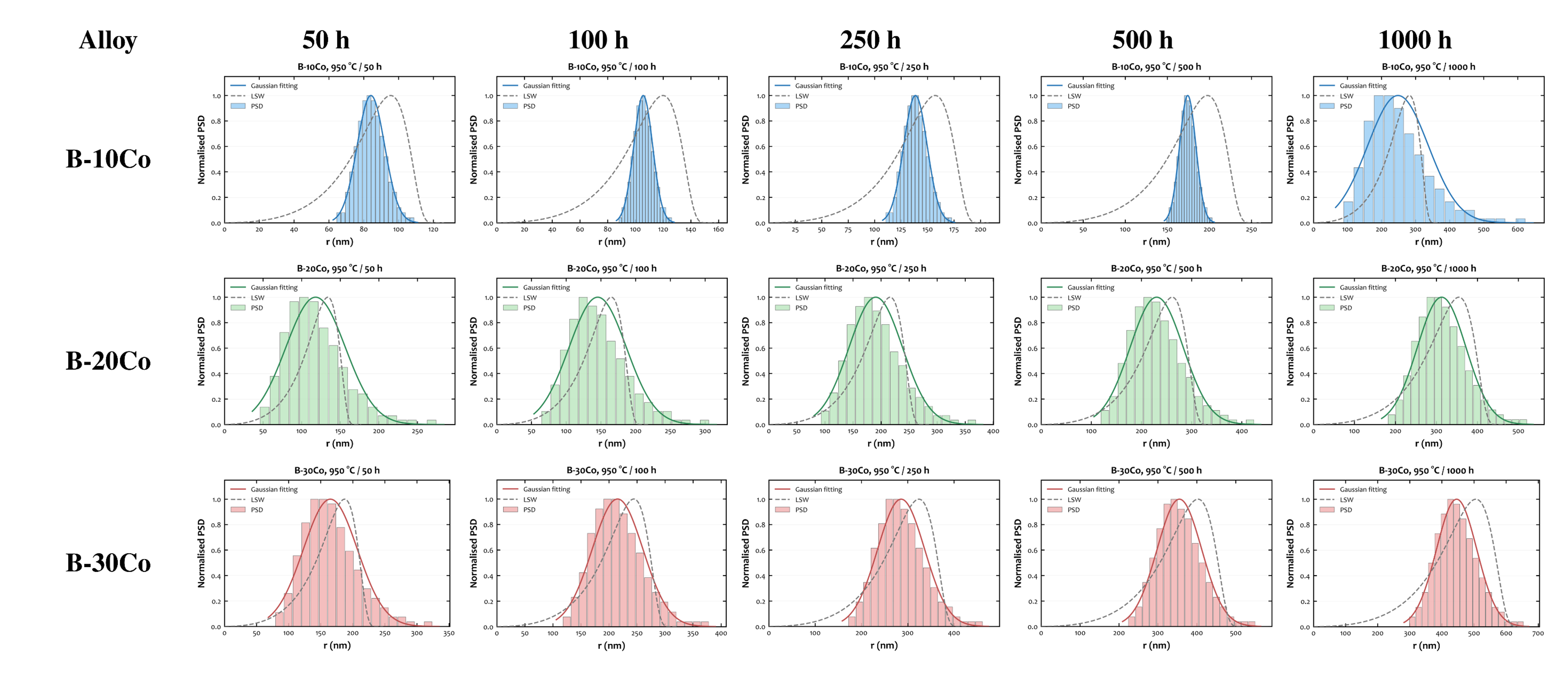}

    \suppcaption{(ii)}
    {Precipitate-size-distribution evolution of the B--10Co, B--20Co, and B--30Co alloys aged at \SI{950}{\degreeCelsius} for 50, 100, 250, 500, and 1000~h. Each plot shows the normalized PSD as a function of precipitate radius, together with the corresponding Gaussian fit and LSW distribution.}

    \phantomsection
    \label{fig:S3_ii}
\end{figure}

% ------------------------- Figure S3(iii) -----------------

\begin{figure}[!htbp]
    \centering

    \includegraphics[
        width=\linewidth,
        height=0.82\textheight,
        keepaspectratio
    ]{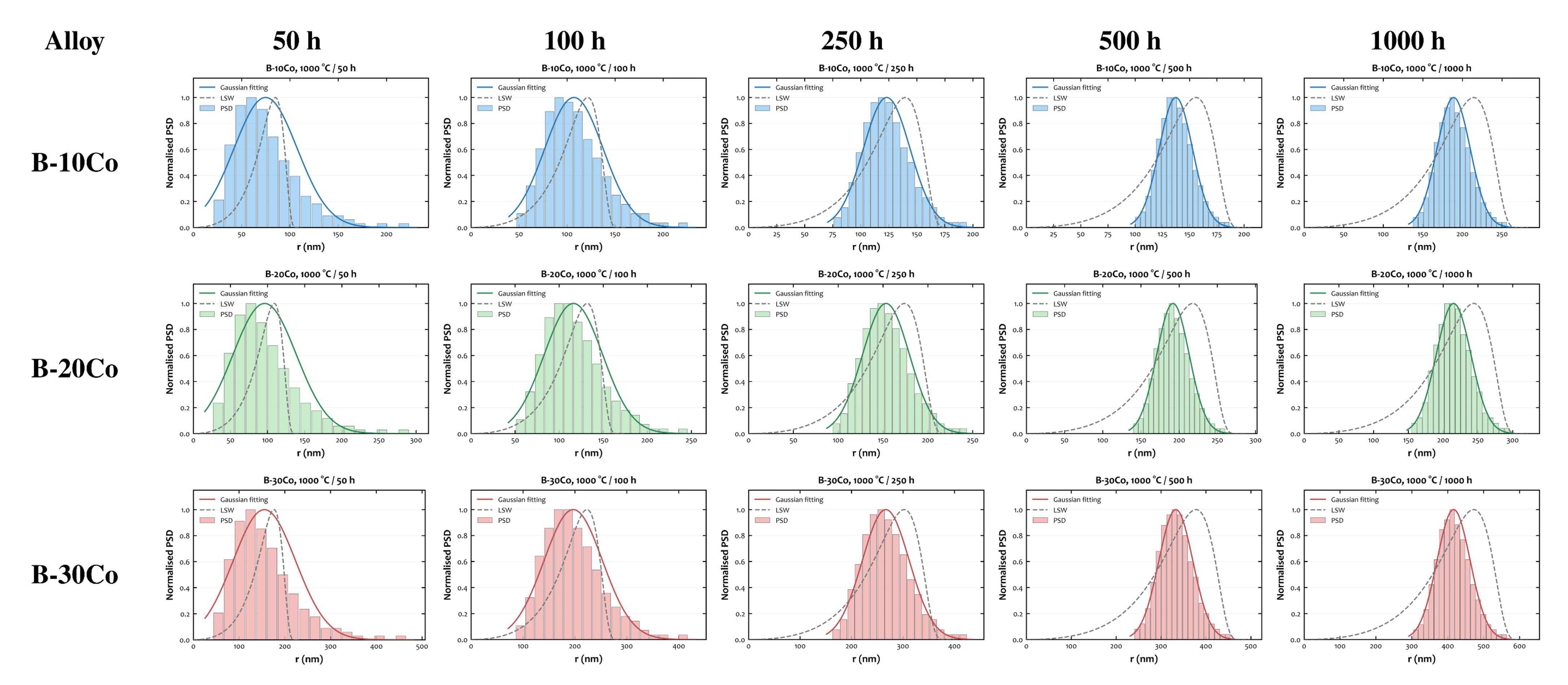}

    \suppcaption{(iii)}
    {Precipitate-size-distribution evolution of the B--10Co, B--20Co, and B--30Co alloys aged at \SI{1000}{\degreeCelsius} for 50, 100, 250, 500, and 1000~h. Each plot shows the normalized PSD as a function of precipitate radius, together with the corresponding Gaussian fit and LSW distribution.}

    \phantomsection
    \label{fig:S3_iii}
\end{figure}

% ------------------------- Figure S3(iv) ------------------

\begin{figure}[!htbp]
    \centering

    \includegraphics[
        width=\linewidth,
        height=0.82\textheight,
        keepaspectratio
    ]{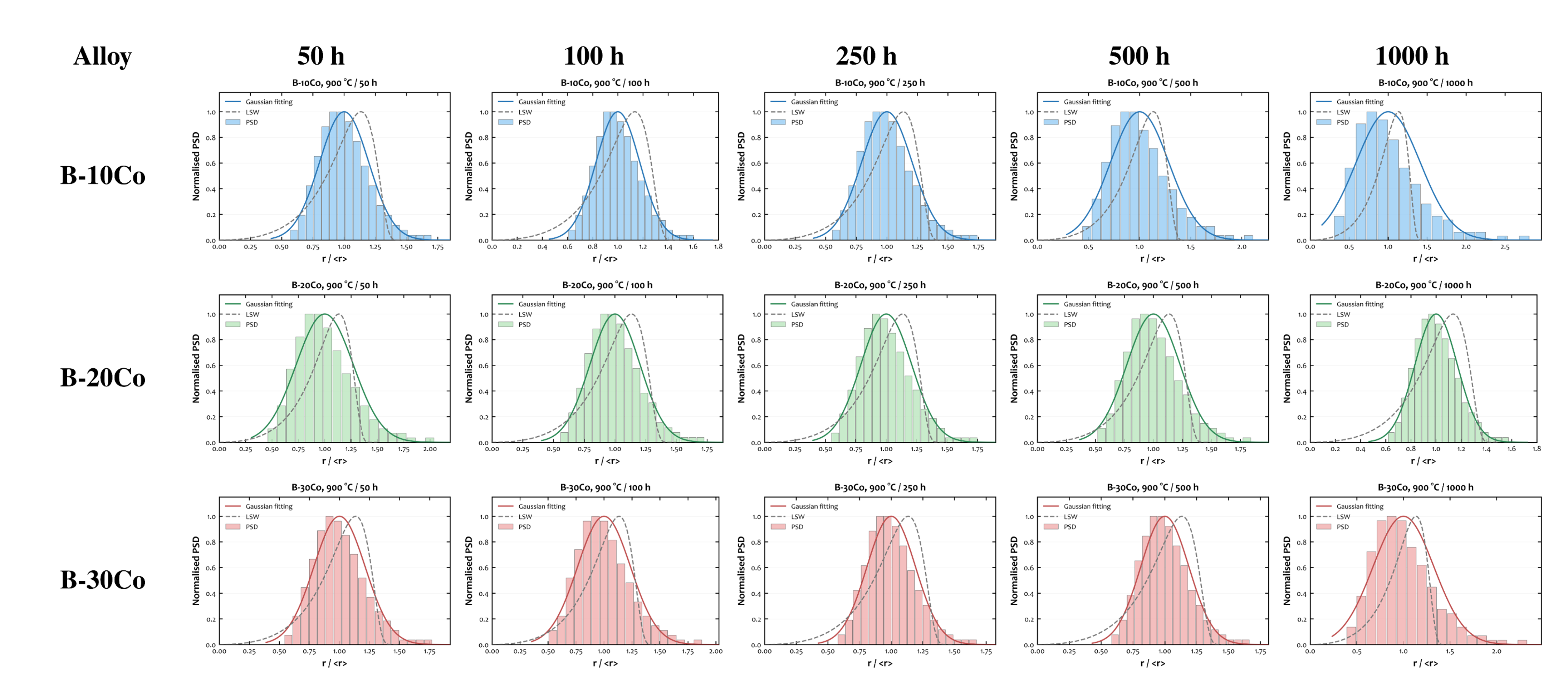}

    \suppcaption{(iv)}
    {Normalized precipitate-size-distribution evolution of the B--10Co, B--20Co, and B--30Co alloys aged at \SI{900}{\degreeCelsius} for 50, 100, 250, 500, and 1000~h. Each plot shows the PSD as a function of normalized precipitate radius, $r/\langle r\rangle$, together with the corresponding Gaussian fit and LSW distribution.}

    \phantomsection
    \label{fig:S3_iv}
\end{figure}

% ------------------------- Figure S3(v) -------------------

\begin{figure}[!htbp]
    \centering

    \includegraphics[
        width=\linewidth,
        height=0.82\textheight,
        keepaspectratio
    ]{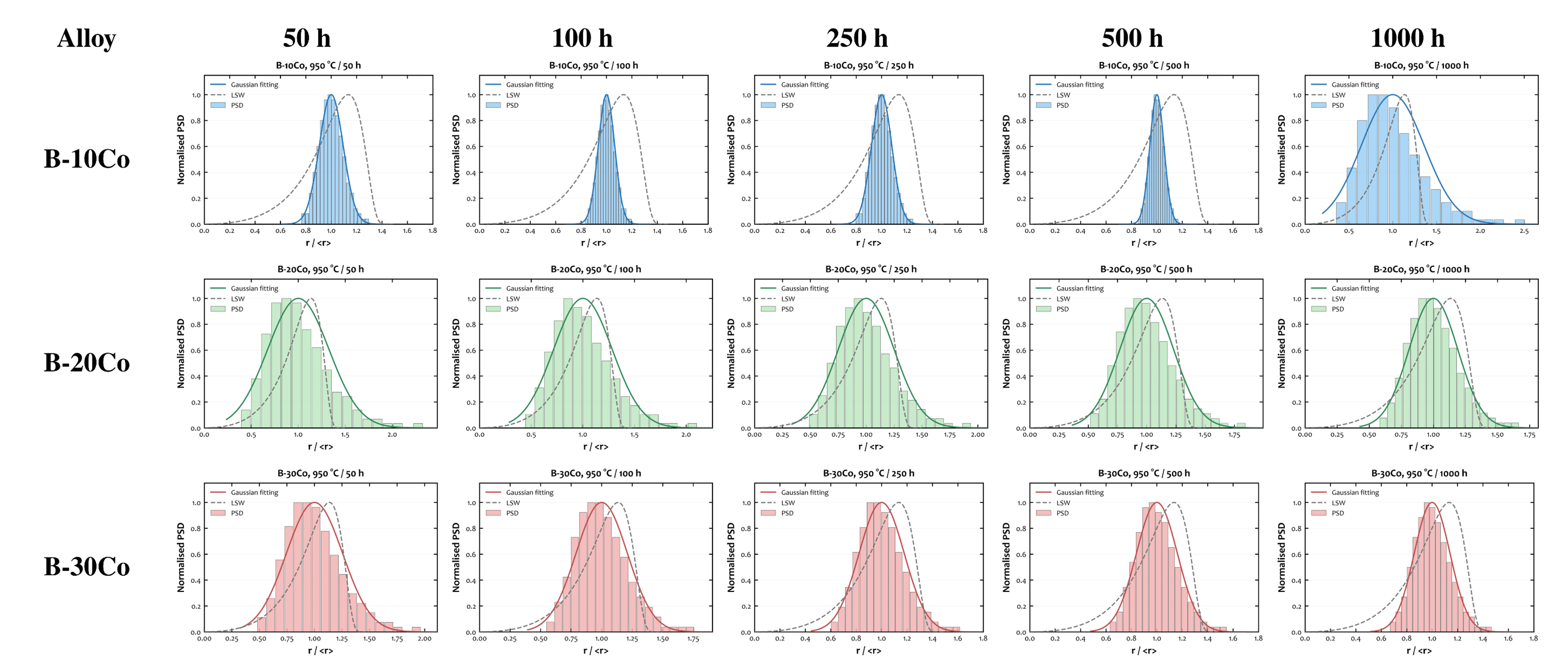}

    \suppcaption{(v)}
    {Normalized precipitate-size-distribution evolution of the B--10Co, B--20Co, and B--30Co alloys aged at \SI{950}{\degreeCelsius} for 50, 100, 250, 500, and 1000~h. Each plot shows the PSD as a function of normalized precipitate radius, $r/\langle r\rangle$, together with the corresponding Gaussian fit and LSW distribution.}

    \phantomsection
    \label{fig:S3_v}
\end{figure}

% ------------------------- Figure S3(vi) ------------------

\begin{figure}[!htbp]
    \centering

    \includegraphics[
        width=\linewidth,
        height=0.82\textheight,
        keepaspectratio
    ]{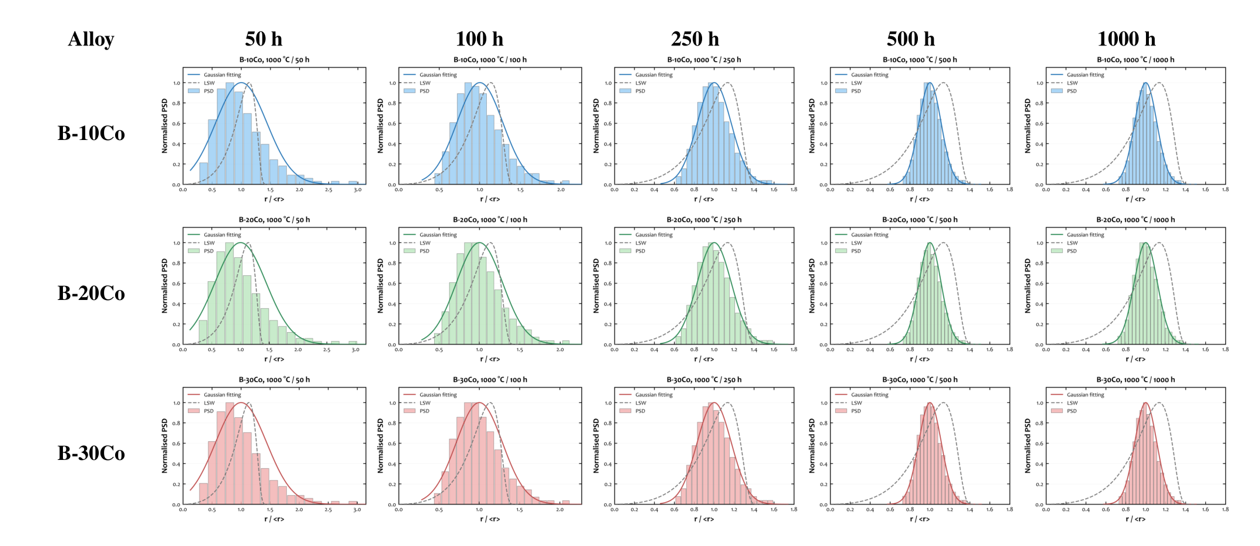}

    \suppcaption{(vi)}
    {Normalized precipitate-size-distribution evolution of the B--10Co, B--20Co, and B--30Co alloys aged at \SI{1000}{\degreeCelsius} for 50, 100, 250, 500, and 1000~h. Each plot shows the PSD as a function of normalized precipitate radius, $r/\langle r\rangle$, together with the corresponding Gaussian fit and LSW distribution.}

    \phantomsection
    \label{fig:S3_vi}
\end{figure}

\end{landscape}

% =========================================================
\clearpage

% BIBLIOGRAPHY
% Materials Science and Engineering: A
% =========================================================

\bibliographystyle{elsarticle-num}
\bibliography{references}

\end{document}